\documentclass[a4paper,11pt]{article}
\usepackage[toc,page]{appendix}
\usepackage{graphicx}
\usepackage{colortbl}
\usepackage{amsmath}
\usepackage{subfigure}
\usepackage{mathtools}
\usepackage[utf8]{inputenc}
\usepackage{float}
\usepackage[normalem]{ulem}
\usepackage{epstopdf}
\usepackage{amsfonts,amsmath,amssymb}
\usepackage{hyperref}

\usepackage{epsfig,multicol,bbm}
\usepackage{color}
\usepackage[dvipsnames]{xcolor}

\definecolor{darkblue}{rgb}{0.0, 0.0, 0.55}
\definecolor{grey}{rgb}{0.57, 0.64, 0.69}
\definecolor{lightbrown}{rgb}{0.71, 0.4, 0.11}

\date{}
\newcommand{\be}{\begin{equation}}
\newcommand{\ee}{\end{equation}}
\usepackage{pgfplots}
\usepgfplotslibrary{fillbetween}
\pgfplotsset{compat=1.18}

\usepackage{pgfplots}
\usepgfplotslibrary{fillbetween}
\pgfplotsset{compat=1.18}

\newcommand\fverb{\setbox\pippobox=\hbox\bgroup\verb}
\newcommand\fverbit{\egroup\item[\fbox{\unhbox\pippobox}]}

\newbox\pippobox
\begin{document}
\title{\bf Generalized Polytropic Regular Black Holes in Arbitrary Dimensions}
\author{Seyed Naseh Sajadi$^{1}$\thanks{Electronic address: naseh.sajadi@gmail.com}\,,\,
Supakchai Ponglertsakul$^{2}$\thanks{Electronic address: supakchai.p@gmail.com}\,,\,
Petarpa Boonserm$^{1}$\thanks{Electronic address: petarpa.boonserm@gmail.com}\,,\,\\
Orlando~Luongo$^{3}$\thanks{Electronic address: orlando.luongo@unicam.it}\,,\,
Hernando Quevedo$^{4}$\thanks{Electronic address: quevedo@nucleares.unam.mx}\,.\,
\\
\small $^1$Department of Mathematics and Computer Science,
Faculty of Science, Chulalongkorn University, \\
\small Bangkok 10330, Thailand,\\
\small $^2$Strong Gravity Group, Department of Physics, Faculty of Science, Silpakorn University,\\ \small Nakhon Pathom 73000, Thailand,\\
\small $^3$Universit\`a di Camerino, Via Madonna delle Carceri 9, 62032 Camerino, Italy,\\
\small $^4$Instituto de Ciencias Nucleares, Universidad Nacional Aut\'onoma de M\'exico, Mexico.\\
}

\maketitle
\begin{abstract}

We investigate static, spherically symmetric regular black holes with anti-de Sitter (AdS) asymptotics in arbitrary spacetime dimensions. They are solutions of Einstein gravity, sourced by an anisotropic fluid whose radial pressure corresponds to vacuum energy, while the tangential pressure satisfies a generalized polytropic equation of state. By solving the Einstein field equations, we derive a generic class of asymptotically AdS black hole solutions and determine the conditions required for spacetime regularity. We then investigate the dynamical formation of these regular black holes within the thin-shell formalism, assuming a linear barotropic equation of state for the shell matter. 
Next, we study the thermodynamics of the regular AdS black holes in arbitrary dimensions by verifying the first law of black hole thermodynamics and the corresponding Smarr relation. We analyze the thermodynamic stability and phase structure of solutions in four, five, and six spacetime dimensions, demonstrating the existence of dimension-dependent phase transitions.

\end{abstract}

\section{Introduction}

Spacetime singularities are among the most important unresolved issues in classical general relativity. According to the singularity theorems, gravitational collapse inevitably leads to singularities under suitable conditions on the matter content and spacetime geometry \cite{Penrose:1964wq,Hawking:1970zq}. Since singularities show that classical gravity breaks down, regular black holes provide a possible solution. They behave like ordinary black holes but do not contain singularities, and all curvature quantities remain finite everywhere.

The first regular black hole solution was introduced by Bardeen \cite{Bardeen:1968}, and since then many regular black hole models have been proposed based on nonlinear electrodynamics \cite{AyonBeato:1998ub,AyonBeato:2000zs,Bronnikov:2022ofk,Bronnikov:2000vy,Sajadi:2017glu,Hendi:2020knv,Bokulic:2026fkz}, modified gravity theories \cite{Bueno:2024dgm,Lobo:2006ue,DeFelice:2026qom,Bueno:2026oyg,Tangphati:2023xnw},  effective quantum gravity corrections \cite{Hayward:2005gi,Sucu:2025pce,Sucu:2026sgc} and dark energy scenarios \cite{adj1,adj2}. 

An alternative approach is to construct regular black holes within Einstein gravity by employing anisotropic matter sources with suitable equations of state \cite{Dymnikova:1992ux,Cheng:2018vux,Sajadi:2023ybm,Sajadi:2025prp}. These models provide a useful framework for understanding how the properties of matter can prevent the formation of curvature singularities.

On the other hand, higher-dimensional black holes have attracted significant attention due to their relevance in string theory, higher-curvature gravity, and holographic theories \cite{Emparan:2008eg}. Extending regular black hole solutions to arbitrary dimensions allows one to explore how dimensionality affects the conditions required for singularity avoidance and the properties of compact objects. However, most existing regular black hole solutions are constructed from specific matter models, and a general framework based on physically motivated equations of state remains an active area of research.

In addition to their static construction, understanding the dynamical formation of regular black holes is an important open problem. The thin-shell formalism provides a powerful method to study gravitational collapse and the evolution of matter shells through the junction conditions \cite{Israel:1966rt,Poisson:2004,Bueno:2025zaj,Bueno:2025gjg,Bueno:2024eig,Bueno:2024zsx}. It has been widely applied to investigate black hole formation, wormholes, and nonsingular compact objects. Nevertheless, the conditions under which regular black holes can emerge dynamically from collapse remain insufficiently explored.

In this work, we construct a class of static, spherically symmetric regular black holes in arbitrary spacetime dimensions within Einstein gravity, sourced by an anisotropic fluid. The radial pressure corresponds to vacuum energy, while the tangential pressure satisfies a generalized polytropic equation of state. We derive the exact solutions and the conditions for spacetime regularity. Using the thin-shell formalism with a linear barotropic equation of state, we also study the dynamical formation of these black holes and show that the collapsing shell either forms a regular black hole or undergoes a nonsingular bounce through the regular core.  We then study the thermodynamics of asymptotically AdS regular black holes, verify the first law of black hole thermodynamics and the Smarr relation, and analyze the phase transitions and thermodynamic stability in four, five, and six spacetime dimensions.

The paper is organized as follows. In Section~\ref{sec2}, we construct the $D$-dimensional static regular black hole solutions and derive the conditions for spacetime regularity. In Section~\ref{shellcolapse}, we investigate the gravitational collapse of a thin shell and determine the conditions for the formation of a regular black hole assuming a linear barotropic equation of state. In Section~\ref{thermosection}, we study the thermodynamics, phase transitions, and thermodynamic stability of the black holes. Finally, in Section~\ref{secconc}, we summarize our results and discuss possible future directions.

\section{Regular $D$-dimensional black hole solutions}\label{sec2}

For a $ D$-dimensional spacetime with cosmological constant $\Lambda$, the Einstein field equations are
\begin{equation}
    \mathcal{R}_{\mu\nu}-\dfrac{1}{2}g_{\mu\nu}\mathcal{R}+\Lambda g_{\mu\nu}=8\pi T_{\mu\nu},
\end{equation}
here, the anisotropic energy-momentum tensor in D-dimensional spacetime is 
\begin{equation}
T^{\mu}{}_{\nu}=\mathrm{diag}\left(-\rho,\,P_r,\,P_t,\,P_t,\,\ldots,\,P_t\right).
\end{equation}
For the static spherically symmetric metric in $D$ dimensions,
\begin{equation}\label{stspBH}
ds^2=-f(r)dt^2+\dfrac{dr^2}{f(r)}+r^2d\Omega_{D-2}^2,
\end{equation}
the nonvanishing components of the Einstein equations are:
\begin{align}
8\pi T^{t}{}_{t}&=\frac{D-2}{2r^2}\left[rf'+(D-3)(f-1)\right]+\Lambda,
\\[1ex]8\pi T^{r}{}_{r}&=\frac{D-2}{2r^2}
\left[rf'+(D-3)(f-1)\right]+\Lambda,\\[1ex]8\pi T^{\theta}{}_{\theta}&=\frac{1}{2r^2}\left[r^2f''+2(D-3)rf'+(D-3)(D-4)(f-1)\right]+\Lambda,
\end{align}
where $f'=df/dr$. For an anisotropic fluid in a static, spherically symmetric spacetime in $D$ dimensions, the generalized Tolman-Oppenheimer-Volkoff (TOV) equation follows from the conservation law ($\nabla_{\mu}T^{\mu}_{\ r}=0$) is given by
\begin{equation}\label{eqtov}
\frac{dP_r}{dr}=-\frac{\rho+P_r}{2}\frac{f'(r)}{f(r)}+\frac{D-2}{r}\left(P_t-P_r\right).
\end{equation}
To construct a class of regular black hole solutions, we adopt the following barotropic equation(s) of state (EoS) \cite{Riazi:2015hga}:
\begin{align}
P_{r}= \omega \rho +\bar{\omega}\dfrac{\rho^{n}}{\rho_{0}^{n-1}},\qquad
P_{t}=\omega_{1}\rho+\omega_{2}\dfrac{\rho^{n}}{\rho_{0}^{n-1}},\label{eos2}
\end{align}
here $\rho_{0}>0$ is the central density, $\omega, \bar{\omega}, \omega_{1}$ and $\omega_{2}$, are the dimensionless EoS parameters.\\
To obtain solution \eqref{stspBH} in Einstein gravity,  we require $\omega=-1$ and $\bar{\omega}=0$. Consequently, the anisotropic fluid is characterized by the following pressure components:
\begin{equation}\label{eqEOS1}
P_{r}=-\rho,\qquad
P_{t}=\omega_{1}\rho+\omega_{2}\frac{\rho^{n}}{\rho_{0}^{\,n-1}}.
\end{equation}
 Plugging the above assumptions into \eqref{eqtov}, one obtains $\rho$,
 \begin{equation}\label{eqqrho}
	\rho(r)=\dfrac{\rho_{0}}{\left[C_{1}\left(\rho_{0}r^{(\omega_{1}+1)(D-2)}\right)^{n-1}+1\right]^{\frac{1}{n-1}}},
\end{equation}
where $C_{1}$ is an integration constant with dimensions $[C_{1}]=L^{(n-1)[2-(\omega_{1}+1)(D-2)]}$ and $\omega_{2}=-\omega_{1}-1$. Assuming a positive energy density, $\rho>0$, the null and weak energy conditions are satisfied provided $\omega_{1}\ge -1$,
while the dominant energy condition requires $-1\le\omega_{1}\le1.$
The strong energy condition is violated near the center, where the equation of state approaches $P_{r}=P_{t}=-\rho.$

By substituting the above energy density into the Einstein field equations and performing the integration\footnote{We use the incomplete beta function identity
\begin{equation}
B(x;a,b)=\int_{0}^{x} t^{a-1}(1-t)^{b-1}dt
=\frac{x^{a}}{a} {}_2F_1(a,1-b;a+1;x).
\end{equation}
},
we obtain the following metric function:
\begin{multline}\label{eqmainmetric}
f(r)=1-\frac{2\Lambda r^{2}}{(D-1)(D-2)}-\frac{\mu}{r^{D-3}}\\
-\frac{16\pi\rho_{0}r^{2}}{(D-1)(D-2)}\,{}_2F_1\!\Biggl(\frac{1}{n-1},\frac{D-1}{(w_{1}+1)(n-1)(D-2)};\\
1+\frac{D-1}{(w_{1}+1)(n-1)(D-2)};-C_{1}\rho_{0}^{\,n-1}r^{(w_{1}+1)(n-1)(D-2)}\Biggr),
\end{multline}
where ${}_2F_1$ is hypergeometric function. Since we are interested in regular black hole solutions, we set the integration constant $\mu=0$ to eliminate the Schwarzschild-like singular contribution.
The total gravitational mass of the fluid distribution is given by

\begin{equation}\label{eqmass}
\begin{aligned}
M&=\Omega_{D-2}\int_{0}^{\infty} r^{D-2}\rho(r)dr=\frac{\Omega_{D-2}\rho_{0}}{(\omega_{1}+1)(D-2)(n-1)}\left(C_{1}\rho_{0}^{\,n-1}\right)^{-\frac{D-1}{(\omega_{1}+1)(D-2)(n-1)}}
\\
&\qquad\times
{B}\!\left(\frac{D-1}{(\omega_{1}+1)(D-2)(n-1)},\,\frac{1}{n-1}-\frac{D-1}{(\omega_{1}+1)(D-2)(n-1)}\right),
\end{aligned}
\end{equation}
where $B(x,y)$ denotes the beta function and
\begin{equation}
\Omega_{D-2}=\frac{2\pi^{\frac{D-1}{2}}}{\Gamma\!\left(\frac{D-1}{2}\right)}
\end{equation}
is the area of the unit $(D-2)$-sphere. The total mass is finite and positive provided that
\begin{equation}\label{eqcond1}
\omega_{1}(D-2)>1, \qquad n>1, \qquad C_{1}>0, \qquad \rho_{0}>0.
\end{equation}
The combined requirements of a finite total mass and the dominant energy condition restrict the parameter $\omega_{1}$ to the range
\begin{equation}\label{eqcond1}
\frac{1}{D-2}<\omega_{1}\le 1.
\end{equation}
This automatically satisfies the null and weak energy conditions as well.
Using the properties of the hypergeometric function\footnote{For $z\to-\infty$ the hypergeometric function is given by
\begin{equation}
{}_2F_1(a,b;c;z)\sim\frac{\Gamma(c)\Gamma(b-a)}{\Gamma(b)\Gamma(c-a)}
(-z)^{-a}+\frac{\Gamma(c)\Gamma(a-b)}{\Gamma(a)\Gamma(c-b)}(-z)^{-b},
\end{equation}
provided that $a-b\notin\mathbb{Z}$. For the special cases with $a-b\in\mathbb{Z}$, the above expression must be replaced by the corresponding integer-difference expansion, which may contain logarithmic terms. In the particular cases considered here, the hypergeometric function reduces 
simply to ${}_2F_1(a,b;a;z)=1/(1-z)^{b}$.}, and under the condition \eqref{eqcond1}, the metric function has the following asymptotic expansion 
\begin{equation}
f(r)=1 -\frac{2\Lambda r^{2}}{(D-1)(D-2)}-\frac{16\pi M}{(D-2)\Omega_{D-2} r^{D-3}}
+\mathcal{O}\left(r^{2-(\omega_{1}+1)(D-2)}\right).
\end{equation}
Therefore, the spacetime is asymptotically (A)dS for $\Lambda\neq0$ and asymptotically flat for $\Lambda=0$.

The near-origin behavior of the metric, obtained using the series expansion of the hypergeometric function, is given by\footnote{For $z\to 0$ the hypergeometric function is given by
\begin{equation}
{}_2F_1(a,b;c;z)=1+\frac{ab}{c}z+O(z^2),\qquad \textrm{valid for}\quad |z|<1.
\end{equation}}

\begin{equation}
\begin{aligned}
f(r)=1-\frac{2\Lambda_{\rm eff}r^{2}}{(D-1)(D-2)}+\frac{16\pi C_{1}\rho_{0}^{\,n}\,r^{2+\alpha}}
{(D-2)(n-1)\left(\alpha+D-1\right)}+\mathcal{O}\left(r^{2+2\alpha}\right).
\end{aligned}
\end{equation}

The spacetime possesses a de Sitter core provided that $\Lambda_{\rm eff}>0$, where the effective cosmological constant is given by
\begin{equation}
\Lambda_{\rm eff}=\Lambda+8\pi\rho_{0},\qquad
\alpha=(\omega_{1}+1)(n-1)(D-2)>0.
\end{equation}

We note that since $\omega_1$ and $n$ are continuous parameters, $\alpha$ is not necessarily an integer. Therefore the metric is not necessarily $C^{\infty}$ near the origin for $\alpha>0$.

The Ricci scalar, the Ricci tensor invariant, and the Riemann tensor invariant near the origin are given by
\begin{equation}
\begin{aligned}
\mathcal{R}={}&\frac{2D\Lambda_{\rm eff}}{D-2}-\frac{16\pi C_1\rho_0^n(D+\alpha)}{(D-2)(n-1)}r^\alpha
+\mathcal{O}\left(r^{2\alpha}\right),\\
\mathcal{R}_{\mu\nu}\mathcal{R}^{\mu\nu}={}&\frac{4D\Lambda_{\rm eff}^{2}}{(D-2)^2}-\frac{64\pi C_1\rho_0^n\Lambda_{\rm eff}(D+\alpha)}
{(D-2)^2(n-1)}r^\alpha+\mathcal{O}\left(r^{2\alpha}\right),\\
\mathcal{R}_{\mu\nu\rho\sigma}\mathcal{R}^{\mu\nu\rho\sigma}={}&\frac{8D\Lambda_{\rm eff}^{2}}{(D-1)(D-2)^2}-\frac{128\pi C_1\rho_0^n\Lambda_{\rm eff}(D+\alpha)}
{(D-1)(D-2)^2(n-1)}r^\alpha+\mathcal{O}\left(r^{2\alpha}\right).
\end{aligned}
\end{equation}
Hence, for $\alpha>0$, all the curvature scalars remain finite at the origin. This confirms that the core of the solution is free from curvature singularities.

The metric \eqref{eqmainmetric} represents a generic class of regular black hole solutions in Einstein gravity with a cosmological constant coupled to an anisotropic matter source satisfying the equation of state \eqref{eqEOS1}. The solution is characterized by the three free parameters $D$, $n$, and $\omega_{1}$, which can be chosen to recover previously known regular black hole solutions in the literature. In particular, for $D=3$ and $D=4$, the above results are reduced to those obtained in \cite{Sajadi:2023ybm,Sajadi:2025prp}.\\  

So far, we have presented a $D$-dimensional multi-polytropic regular black hole sourced by anisotropic matter. We then address the following question: Can this type of regular black hole form through the collapse of a thin-shell? In the next section, we study the dynamical formation of a black hole by considering the collapse of a thin shell.

\subsection{Gravitational shell collapse}\label{shellcolapse}

We consider the collapse of a spherical shell of matter with the surface stress energy tensor taking the form $S_{ab}=\sigma u_{a}u_{b}+ph_{ab}$, where $h_{ab}=g_{ab}+u_{a}u_{b}$ is an induce metric on $\Sigma_{D-2}$, $\sigma$ is the surface energy density of matter and $u_{a}$ is its velocity. 
At a given moment of proper time, we set the radius of the shell to $r=R(\tau)$. Inside the shell, $r<R(\tau)$, we take the metric to be $f_{-}$. The exterior of the shell, $r>R(\tau)$, $f_{+}$ is a solution to the field equation in $D$ dimension. Therefore, the spacetime consists of two parts that are joined at the location of the shell,   
\begin{equation}
ds_{\pm}^2=-f_{\pm}(r)dt_{\pm}^2+\dfrac{dr^2}{f_{\pm}(r)}+r^2d\Omega_{D-2}^2.
\end{equation}
 Our goal is to determine the evolution of the shell radius. To this end, we use the junction conditions. Therefore, each side of the hypersurface of the thin-shell is parameterized as $(t_{\pm},r)=(T_{\pm}(\tau), R(\tau))$. The first junction condition requires that the metric be continuous across the shell. The induced metric on the shell is
\begin{equation}
ds^{2}_{\Sigma}=-\left(f(R) \dot{T}_{\pm}^{2}-\dfrac{\dot{R}^{2}}{f_{\pm}(R)}\right)d\tau^2+R(\tau)^2d\Omega_{D-2}^2,
\end{equation}
where $\dot{T}=dT/d\tau$. By demanding continuity, we find that
\begin{equation}
f_{\pm}(R)\dot{T}_{\pm}=\pm\sqrt{\dot{R}^2+f_{\pm}(R)}\equiv \beta_{\pm}.
\end{equation} 
The second junction condition for a thin shell with surface stress energy tensor is given by 
\begin{equation}
[K_{ab}]-h_{ab}[K]=-8\pi S_{ab},
\end{equation}
where $K_{ab}=n_{\alpha;\beta}e^{\alpha}_{\ a}e^{\beta}_{\ b}$ is the extrinsic curvature, $K=n^{\alpha}_{\ ;\alpha}$ its trace and $h_{ab}$ is the induced metric on $\Sigma$.
To apply the second junction conditions we need to determine $n^{\alpha}$ and $u^{\alpha}$ as
\begin{align}
u_{\pm}^{\alpha}=\dot{T}_{\pm}\delta^{\alpha}_{t}+\dot{R}\delta^{\alpha}_{r},\;\;\;\;n^{\alpha}_{\pm}=-\dot{R}\delta^{\alpha}_{t}+\dot{T}_{\pm}\delta^{\alpha}_{r}.
\end{align}
The straightforward computation reveals that 
\begin{align}
K^{\pm}_{\tau\tau}=-\dfrac{\dot{\beta}_{\pm}}{\dot{R}},\;\;\;\;\;K^{\pm}_{\phi\phi}=\beta_{\pm}R,\;\;\;\;\beta_{\pm}=\epsilon_\pm\sqrt{\dot{R}^2+f_{\pm}},
\end{align}
where
\begin{equation}
\epsilon_\pm=\operatorname{sign}\left(n^\mu_\pm\partial_\mu r\right)=\pm 1.
\end{equation}
The sign of $\epsilon$ indicates whether the normal points toward increasing or decreasing $r$, respectively. For the configuration considered here, we choose the normal to point from the interior to the exterior, so that $\epsilon_\pm=+1$. This choice remains valid throughout the collapse, since $\beta_\pm$ do not change sign.

Using the above components, the surface stress-energy tensor is given as 
\begin{align}
8\pi S^{\tau}_{\tau} &= -\dfrac{(D-2)}{R}\left(\beta_{-}-\beta_{+}\right),\label{eq30eq}\\
8\pi S^{\theta}_{\theta}
&= \dfrac{1}{R^{D-3}}\frac{d}{dR}
\left[
R^{D-3}\left(\beta_{+}-\beta_{-}\right)
\right].
\label{eq8}
\end{align}
It is easy to rewrite the above equations using $S^{\tau}_{\tau}=-\sigma$ and 
$S^{\theta}_{\theta}=p$ as follows
\begin{equation}\label{eqcon}
    \dfrac{d(\sigma R^{D-2})}{dR}+(D-2)pR^{D-3}=0.
\end{equation}
In the following, we analyze the collapse of the thin shell with a linear EoS.

\subsection{Collapse of linear barotropic equation of state}

We consider a perfect fluid obeying the linear barotropic EoS
\begin{equation}
     p=\zeta\,\sigma.
\end{equation}
Using \eqref{eqcon} and the above EoS, $\sigma$ can be obtained as follows
\begin{equation}\label{eqqrho}
	\sigma(R)=\sigma_{0}R^{-(\zeta+1)(D-2)}.
\end{equation}
Assuming a positive surface energy density $\sigma_{0}>0$, the energy conditions impose constraints on the parameter $\zeta$. The null and weak energy conditions are satisfied provided $\zeta\geq -1$, while the dominant energy condition requires $-1 \leq \zeta \leq 1$.
Moreover, the strong energy condition implies $\zeta\geq-\frac{1}{D-2}$.
Therefore, for $-1\leq\zeta<-\frac{1}{D-2}$, the shell satisfies the null, weak, and dominant energy conditions, while violating the strong energy condition. In this case, the shell consists of matter with negative pressure (tension).

By inserting \eqref{eqqrho} into the equation \eqref{eq30eq}, we get
\begin{equation}\label{eqshelldynamics}
    \dot{R}^2+V_{\rm eff}(R)=0,
\end{equation}
here
\begin{align}\label{eqVeff}
    V_{\rm eff}=\frac{f_{-}+f_{+}}{2}-\frac{(f_{-}-f_{+})^{2}(D-2)^{2}}{256\pi^{2}\sigma^{2}R^{2}}-\frac{16\pi^{2}R^{2}\sigma^{2}}{(D-2)^{2}}.
\end{align}
We now consider the following spacetime metrics for the interior and exterior of the shell 
\begin{equation}\label{first}
	f_{+}=f(R), \;\;\;\;\; f_{-}=1-\frac{2\Lambda R^{2}}{(D-1)(D-2)}.
\end{equation}
For a regular black hole, the metric functions near the origin behave as
\begin{align}
f_+(R) = 1-\frac{2\Lambda_{\rm eff}R^2}{(D-1)(D-2)}+\mathcal{O}(R^{2+\alpha}),\qquad f_-(R)=1-\frac{2\Lambda R^2}{(D-1)(D-2)} .
\end{align}
We obtain the near-origin expansion of the effective potential
\begin{equation}
\begin{aligned}
V_{\rm eff}(R) &= \,1-\frac{\Lambda+\Lambda_{\rm eff}}{(D-1)(D-2)}R^2-\frac{(\Lambda_{\rm eff}-\Lambda)^2}
{64\pi^2\sigma_0^2(D-1)^2}R^{2+2(\zeta+1)(D-2)}\nonumber\\
&~~~~-\frac{16\pi^2\sigma_0^2}{(D-2)^2}R^{2-2(\zeta+1)(D-2)}+\cdots .
\end{aligned}
\end{equation}
The dominant behavior of the effective potential near the origin depends on the value of $\zeta$, and is summarized as follows:

\begin{equation}\label{nearoriginresults}
V_{\rm eff}(R)\longrightarrow
\begin{cases}
-\infty,
&
\displaystyle \zeta>-\frac{D-3}{D-2},
\\[1.2ex]
\displaystyle
1-\frac{16\pi^2\sigma_0^2}{(D-2)^2},
&
\displaystyle \zeta=-\frac{D-3}{D-2},
\\[1.2ex]
1,
&
\displaystyle
-\frac{D-1}{D-2}<\zeta<-\frac{D-3}{D-2},
\\[1.2ex]
\displaystyle
1-\frac{\rho_{0}^2}
{\sigma_0^2(D-1)^2},
&
\displaystyle \zeta=-\frac{D-1}{D-2},
\\[1.2ex]
-\infty,
&
\displaystyle \zeta<-\frac{D-1}{D-2}.
\end{cases}
\qquad (R\rightarrow 0).
\end{equation}
For large $R$, the effective potential behaves as
\begin{equation}
\begin{aligned}
V_{\rm eff}
=&\;
1-\frac{2\Lambda R^{2}}{(D-1)(D-2)}
-\frac{8\pi M}{(D-2)\Omega_{D-2}R^{D-3}}
\\
&\;
-\frac{M^{2}}{\Omega_{D-2}^{2}\sigma_{0}^{2}}
R^{2(D-2)\zeta}
-\frac{16\pi^{2}\sigma_{0}^{2}}{(D-2)^{2}}
R^{-2(D-3)-2(D-2)\zeta}
+\cdots .
\end{aligned}
\end{equation}
Therefore, in the asymptotically flat case ($\Lambda=0$), the effective potential asymptotically behaves as
\begin{equation}\label{largerresults}
V_{\rm eff}(R)\rightarrow
\begin{cases}
-\infty,
&
\zeta>0,
\\[1ex]
1-\dfrac{M^{2}}{\Omega_{D-2}^{2}\sigma_{0}^{2}},
&
\zeta=0,
\\[1ex]
1,
&
-\dfrac{D-3}{D-2}<\zeta<0,
\\[1ex]
1-\dfrac{16\pi^{2}\sigma_{0}^{2}}{(D-2)^{2}},
&
\zeta=-\dfrac{D-3}{D-2},
\\[1ex]
-\infty,
&
\zeta<-\dfrac{D-3}{D-2}.
\end{cases}
\qquad (R\rightarrow\infty).
\end{equation}
Combining the near-origin and asymptotic analyses for asymptotically flat spacetime, we find that the effective potential remains finite at both boundaries provided that
\begin{equation}\label{eqcond2}
\zeta = -\dfrac{D-3}{D-2}.
\end{equation}
This condition follows from the specific requirement that the effective potential remain finite at both boundaries and should not be interpreted as a universal condition for the formation of regular black holes.

By substituting \eqref{eqcond2} into \eqref{eqqrho}, we obtain
\begin{equation}
    \sigma(R)=\frac{\sigma_{0}}{R},
\end{equation}
and consequently, the effective potential in \eqref{eqVeff} becomes
\begin{align}\label{eqVeff1}
    V_{\rm eff}
    =\frac{f_{-}+f_{+}}{2}
    -\frac{(f_{-}-f_{+})^{2}(D-2)^{2}}
    {256\pi^{2}\sigma_{0}^{2}}
    -\frac{16\pi^{2}\sigma_{0}^{2}}{(D-2)^{2}}.
\end{align}
The effective potential near the origin $(R\to0)$ takes the form
\begin{equation}
V_{\rm eff}(R) =1-\frac{16\pi^2\sigma_0^2}{(D-2)^2}-\frac{8\pi\rho_{0}}{(D-1)(D-2)}R^2
-\frac{\rho_{0}^2}{\sigma_0^2(D-1)^2}R^4+\mathcal{O}(R^6).
\end{equation}
If $\sigma_0>\frac{D-2}{4\pi}$, then $V_{\rm eff}(0)<0$, and the shell reaches the origin in finite proper time. If 
\begin{equation}
\sigma_0=\frac{D-2}{4\pi},
\end{equation}
then the effective potential vanishes at the origin, $V_{\rm eff}(0)=0$. For $\rho_0>0$, its leading behavior near the origin is given by
\begin{equation}
V_{\rm eff}(R) \sim -\frac{8\pi\rho_{0}} {(D-1)(D-2)}R^2,
\end{equation}
and consequently
\begin{equation}
R(\tau)\sim R_0 \exp\left[ -\sqrt{\frac{8\pi\rho_{0}}{(D-1)(D-2)}}(\tau-\tau_0)\right].
\end{equation}
Thus, in this case, the shell approaches the origin $R=0$ only in the infinite proper time limit. Therefore, combining the results obtained above, we find that, in order to obtain a regular black hole through the gravitational collapse of the thin shell, the shell pressure and energy density must take the form
\begin{equation}
p=-\frac{D-3}{4\pi R},
\qquad
\sigma=\frac{D-2}{4\pi R}.
\end{equation}
Thus for $D>3$, the collapsing shell has a negative pressure and positive density. 

We then study the gravitational collapse of three models of regular black holes in different spacetime dimensions and matter-source parameters in the following subsections.

 \subsection{The case $D=4$}\label{subsecD4}

For the particular choice \(n=\frac{3}{2}\) and \(\omega_{1}=2\), the metric function \eqref{eqmainmetric} is reduced to\footnote{We note that this example, with $\omega_1=2$, does not satisfy the dominant energy condition, since the bound $\omega_1\leq1$ applies only when the dominant energy condition is imposed. We have chosen this particular metric because it corresponds to the Hayward metric considered in \cite{Hayward:2005gi}.} \cite{Sajadi:2025prp}
\begin{equation}\label{eqmetrci55}
f(r)=1-\frac{\Lambda r^{2}}{3}-\frac{8\pi\rho_{0}r^{2}}
{3\left(1+C_{1}\sqrt{\rho_{0}}\,r^{3}\right)},
\end{equation}
here 
\begin{equation}
    C_{1}=\dfrac{4\pi\sqrt{\rho_{0}}}{3M}.
\end{equation}

The metric function is regular at the origin and approaches the Schwarzschild-(A)dS form at large distances. Near the origin, the metric function behaves as 
\begin{equation}
f(r)=1-\frac{(\Lambda+8\pi\rho_{0})}{3}r^{2}+\mathcal{O}(r^{5}),
\end{equation}
which shows that the central region is regular and has a de Sitter-like structure when $\Lambda+8\pi\rho_{0}>0$. At large $r$, the metric function takes the form of an asymptotically Schwarzschild-(A)dS spacetime
\begin{equation}
f(r)=1-\frac{2M}{r}-\frac{\Lambda r^{2}}{3}+\mathcal{O}(r^{-4}).
\end{equation}
The mass function is also regular at the origin and approaches the total mass $M$ at infinity.

\begin{figure}
\centering
\includegraphics[width=0.32\columnwidth]{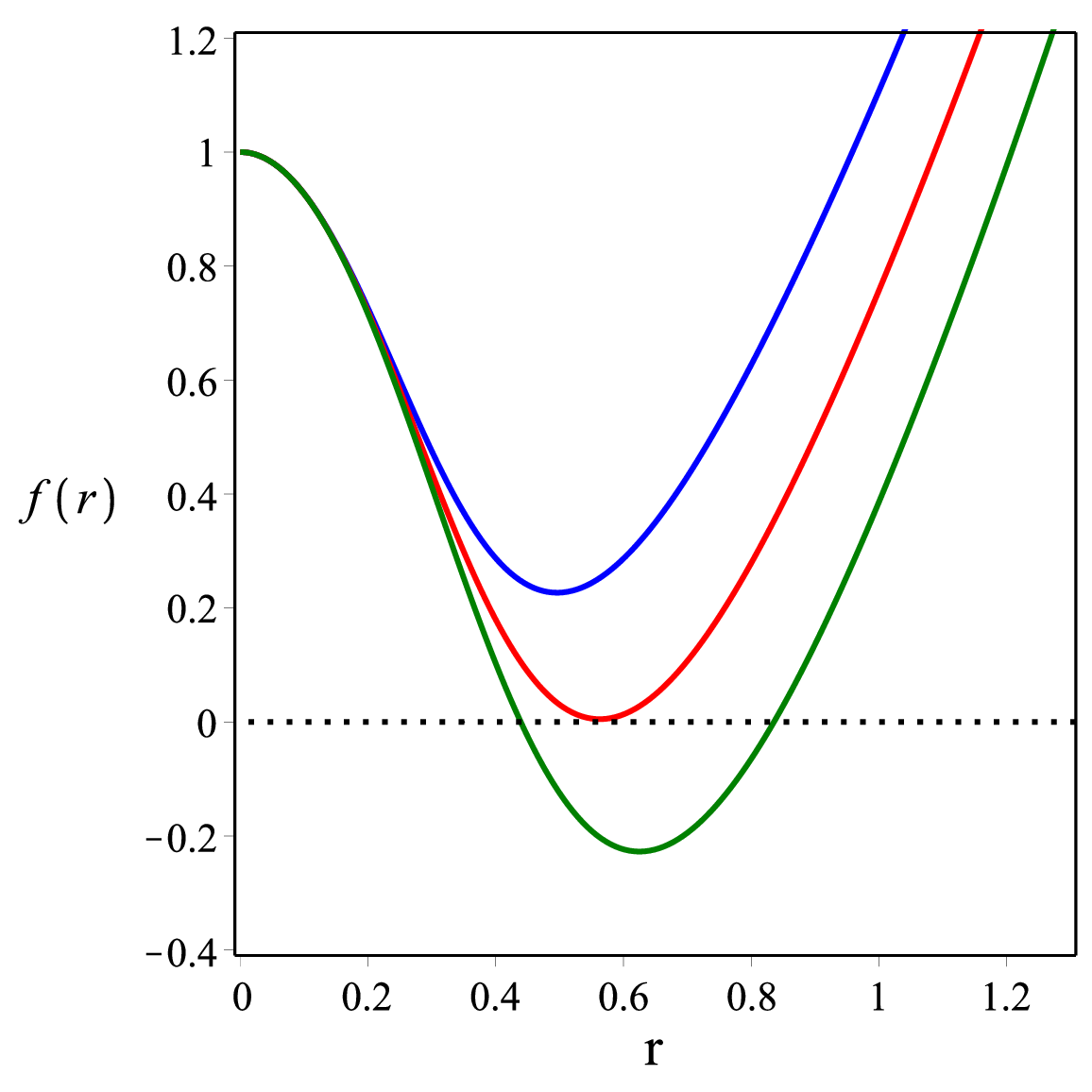}
\includegraphics[width=0.32\columnwidth]{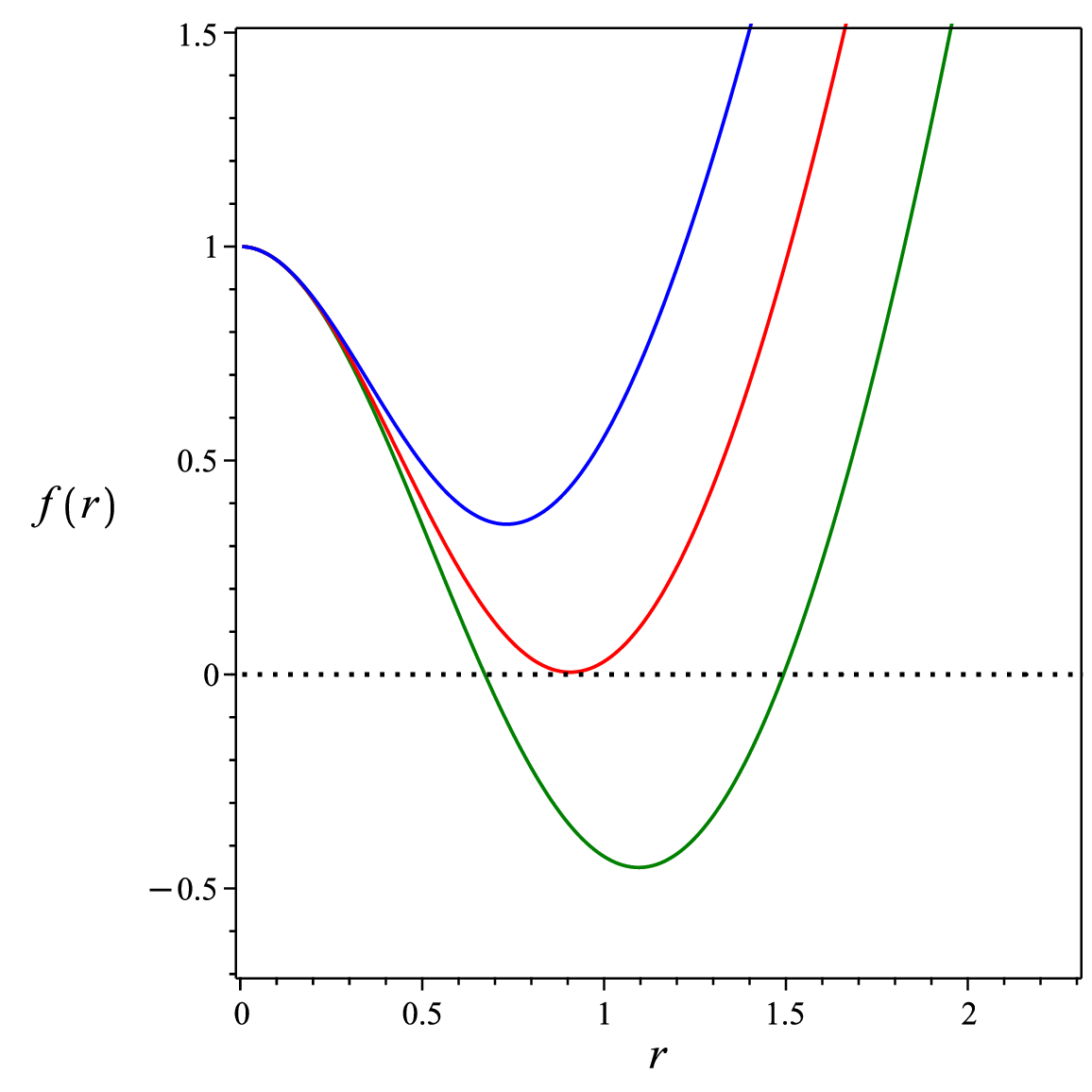}
\includegraphics[width=0.32\columnwidth]{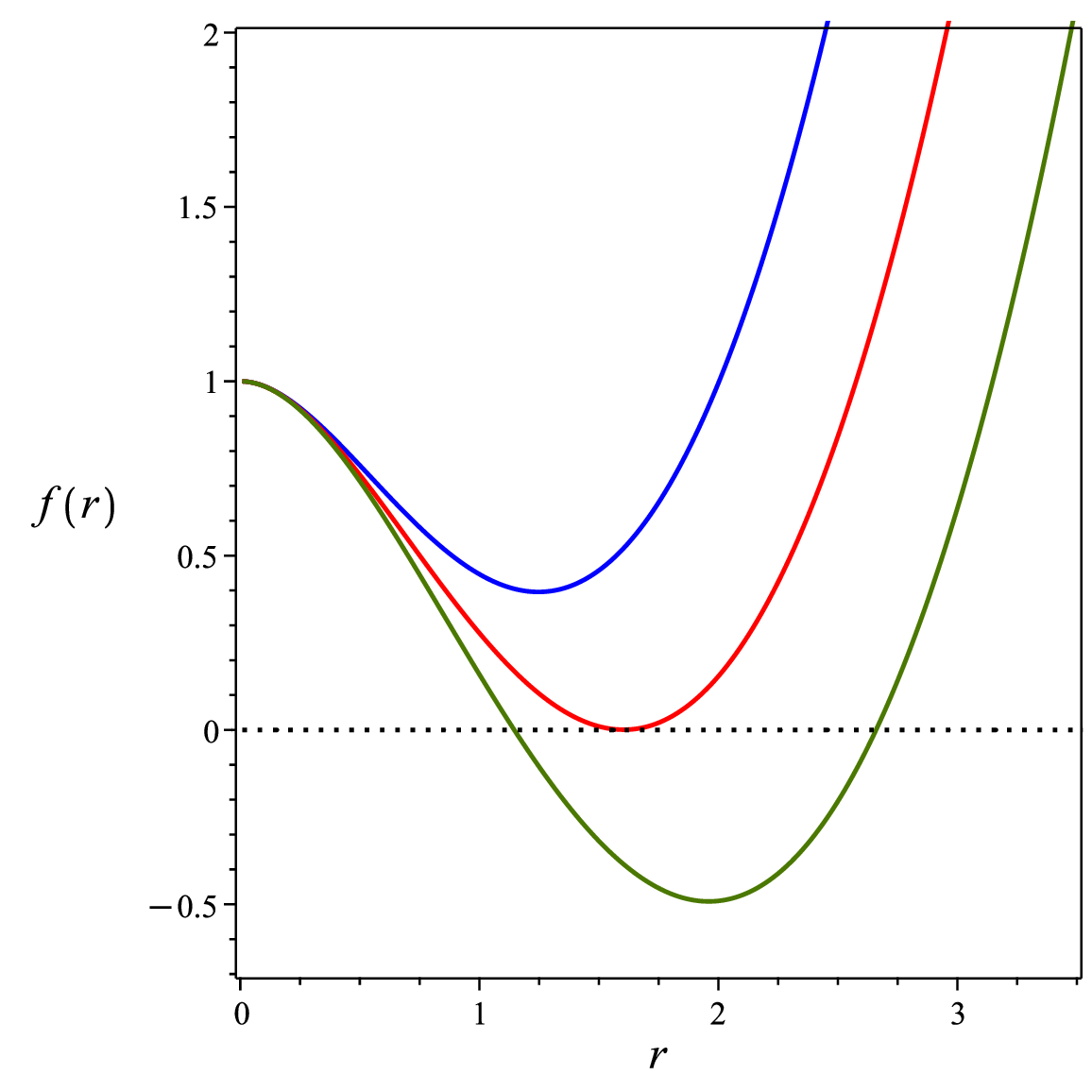}
\caption{Left: The metric function \eqref{eqmetrci55} as a function of the radial coordinate $r$ for $D=4$, and $M={1.0},\;{0.73},\;{0.5}$ (from bottom to top). Middle: The metric function \eqref{metric5d} as a function of the radial coordinate $r$ for $D=5$, and $M={50},\;{23.5},\;{10}$ (from bottom to top). Right: The metric function \eqref{eqmetric6d} as a function of the radial coordinate $r$ for $D=6$, and $M={4788800},\;{1758800},\;{500000}$ (from bottom to top). We have set $\rho_{0}=1$, $\ell=1$.}
\label{fig:metric0}
\end{figure}

The extremal black hole is determined by
\begin{equation}
M_{\rm ext}=\frac{r_h}{12}\left(\Lambda r_h^2-3\right)^2,
\qquad
\rho_{0}^{\rm ext}=-\frac{\left(\Lambda r_h^2-3\right)^2}{8\pi r_h^2\left(\Lambda r_h^2-1\right)}.
\end{equation}
For $M>M_{\rm ext}$, spacetime has two horizons. In the extremal case, $M=M_{\rm ext}$, the two horizons coincide, forming a single degenerate horizon. For $M<M_{\rm ext}$, no event horizon exists and the spacetime describes a horizonless geometry. This behavior is illustrated in the left panel of figure~\ref{fig:metric0}.

Then, following the approach presented in the previous section, we consider the formation of the black hole \eqref{eqmetrci55} through the gravitational collapse of a thin shell.
 The shell equation is given by
\begin{equation}
\dot{R}^2-\frac{4\pi\rho_0R^2 M}{3M+4\pi\rho_0R^3}\left[1+
\frac{\pi\rho_0 R M}{3M+4\pi\rho_0R^3}\right]=0.
\end{equation}
Near the origin, the effective potential can be expanded as
\begin{equation}
V_{\rm eff} =-\frac{4\pi\rho_0}{3}R^2 -\frac{4\pi^2\rho_0^2}{9}R^4 +\mathcal{O}(R^5).
\end{equation}
Thus, the shell radius behaves as
\begin{equation}
R(\tau) \sim R_0 \exp\left[-\sqrt{\frac{4\pi\rho_0}{3}}(\tau-\tau_0)\right].
\end{equation}
This shows that the shell approaches the origin $R=0$ in the infinite proper time.

\begin{figure}
\centering
\includegraphics[width=0.45\columnwidth]{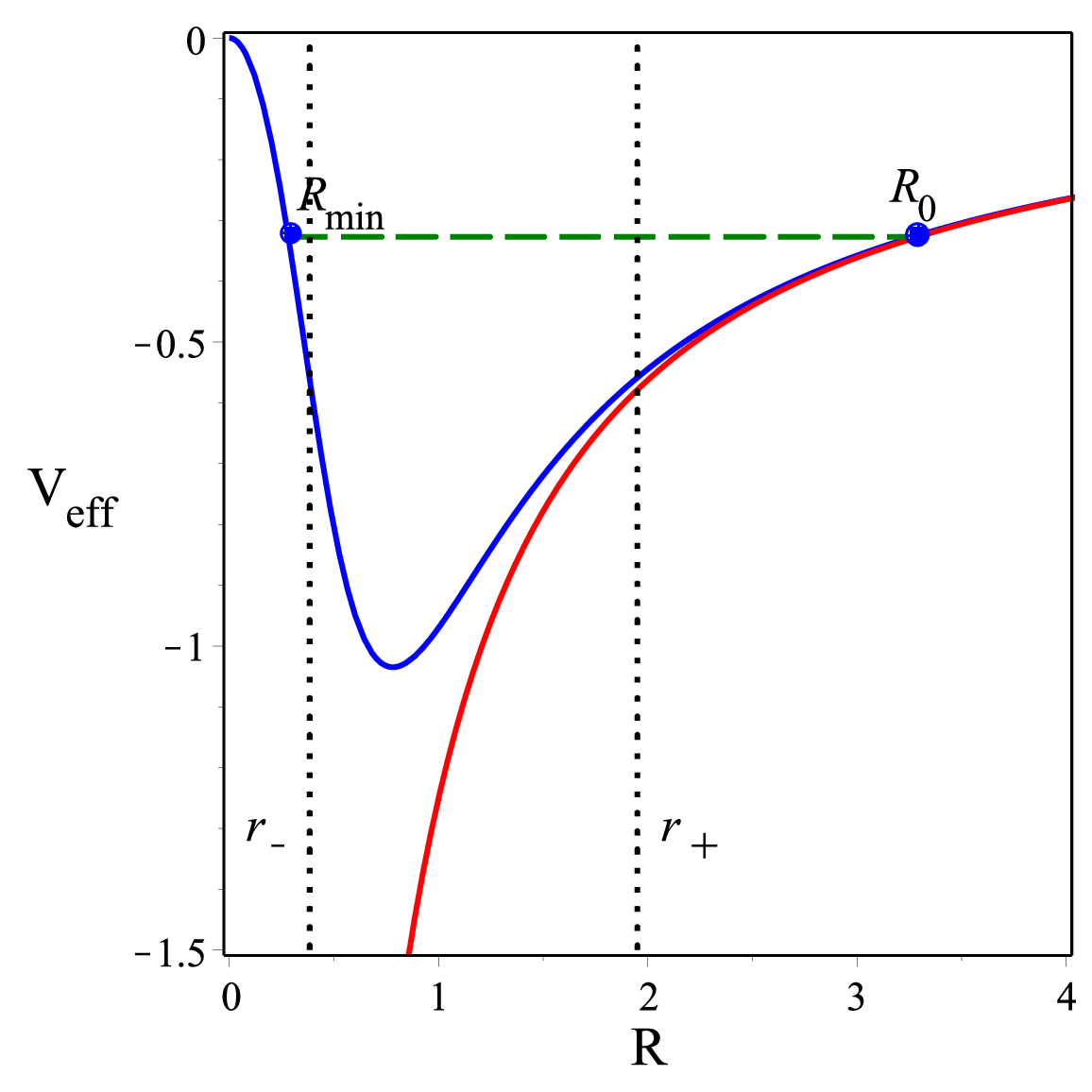}
\includegraphics[width=0.45\columnwidth]{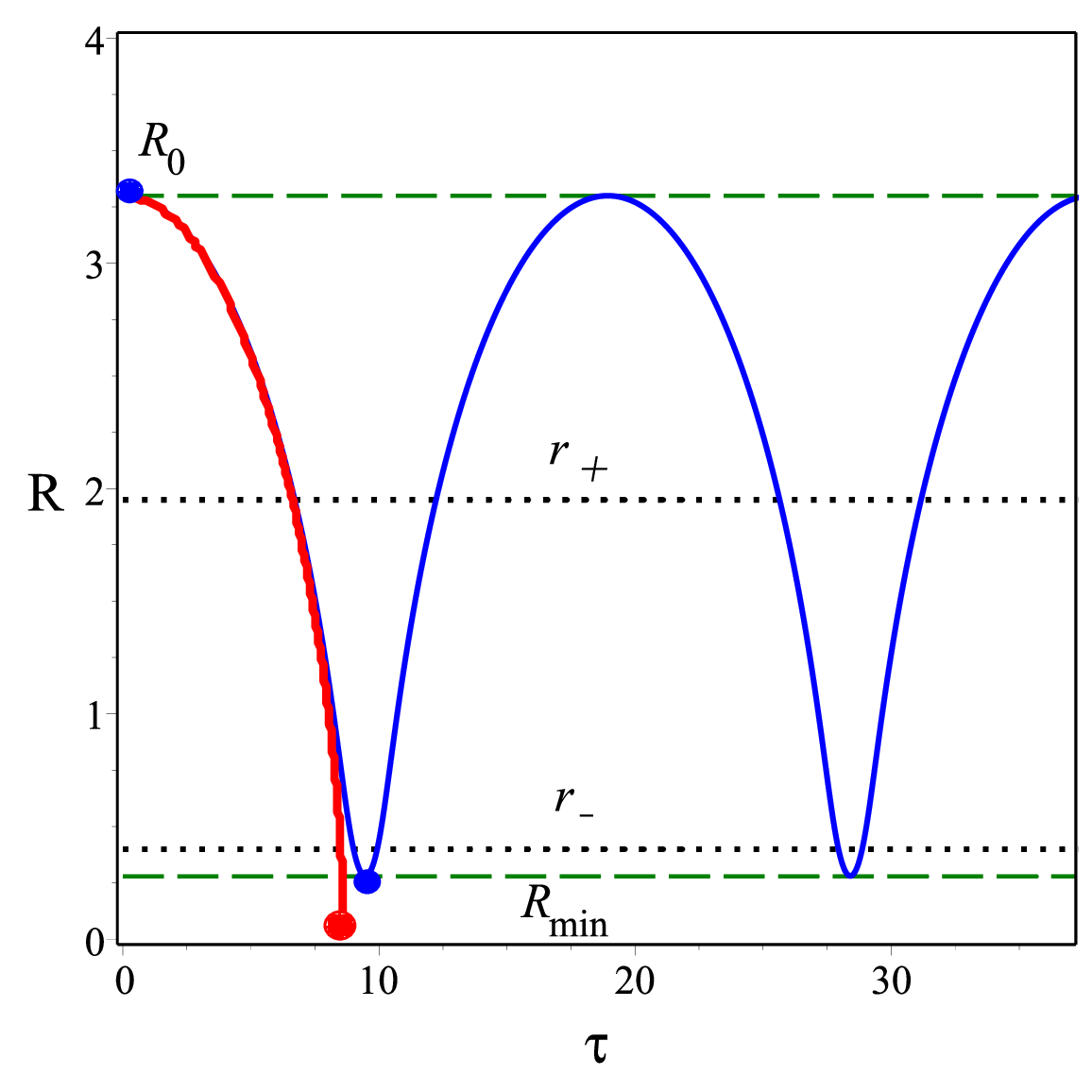}
\caption{Effective potential $V_{\rm eff}$ governing the thin-shell motion (left panel) and the shell radius $R$ as a function of proper time $\tau$ (right panel). The blue curves correspond to the regular black hole, while the red solid curves represent the Schwarzschild black hole. For the regular black hole, the shell collapses from $R_{0}=3.3$, crosses the horizons, reaches a finite minimum radius $R_{\min}=0.28$, and then bounces back. The shell subsequently expands into a new asymptotic region, and the collapse--bounce process repeats whenever the shell returns to $R_{0}$. In contrast, the Schwarzschild shell continues to collapse toward $R=0$, where the singularity is formed. The parameters are fixed as $D=4$, $\Lambda=0$, $\omega_{1}=2$, $n=3/2$, $M=1$, $\rho_{0}=1$, and $\zeta=-1/2$.
}
\label{fig:phasek3}
\end{figure}

Figure~\ref{fig:phasek3} shows the effective potential and the evolution of the thin shell during gravitational collapse. The left panel shows the effective potential $V_{\rm eff}(R)$ for the $4D$ regular black hole (blue curve) and the Schwarzschild black hole (red curve), while the right panel shows the corresponding evolution of the shell radius $R$ with respect to its proper time $\tau$.

The shell initially starts at a finite radius $R_{0}>r_{+}$ and collapses inward. For the regular black hole, the shell crosses the outer and inner horizons, $r_{+}$ and $r_{-}$, respectively. As it moves further inward, the shell reaches a finite minimum radius $R_{\min}<r_{-}$, corresponding to the turning point of the effective potential. At this point, the shell bounces and starts to expand. It then crosses the inner and outer horizons again and moves into a new asymptotic region. The shell can subsequently reach $R_{0}$, leading to a repeated collapse and bounce.

In contrast, for the Schwarzschild black hole, the effective potential does not provide a finite turning point that prevents the shell from reaching the center. As shown by the red curve in the right panel, the shell continues to collapse toward $R=0$, where the Schwarzschild singularity is formed. The comparison between the two panels therefore shows that the regular black hole replaces the singular collapse by a bounce at a finite radius. This behavior demonstrates the nonsingular nature of the regular black hole and shows how the gravitational collapse can lead to a regular black hole rather than a singular one.

\subsection{The case $D=5$}\label{subsecD5}

For the particular choice $n=4/3$ and $\omega_{1}=1$, the metric function takes the following simple form:
\begin{equation}\label{metric5d}
f(r)=1-\frac{\Lambda r^{2}}{6} -\frac{4\pi\rho_{0}r^{2}}
{3\left(1+C_{1}\rho_{0}^{1/3}r^{2}\right)^{2}},
\end{equation}
where the constant $C_{1}$ is related to the mass $M$ by
\begin{equation}
C_{1}^{2}=\frac{\pi^{2}\rho_{0}^{1/3}}{2M}.
\end{equation}
This metric describes a five-dimensional black hole with a matter contribution. In particular, the metric function remains finite at the origin, while its large-$r$ behavior approaches that of a {five-dimensional} Schwarzschild-(A)dS black hole.

The extremal configuration corresponds to the case where the inner and outer horizons coincide. It is therefore obtained by requiring the metric function and its first derivative to vanish simultaneously at the horizon radius $r_h$.
Solving them gives the extremal mass and the corresponding central density as
\begin{align}
M_{\mathrm{ext}} =-\frac{\pi r_h^2}{144}\left(\Lambda r_h^2-6\right)^3,\qquad 
\rho^{\mathrm{ext}}_{0}=-\frac{\left(\Lambda r_h^2-6\right)^3}
{8\pi r_h^2\left(\Lambda r_h^2-3\right)^2}.
\end{align}
These expressions determine the parameter values for which the black hole becomes extremal and provide the boundary between configurations with two horizons and configurations without a horizon. The behavior of the metric function for the non-extremal, extremal, and horizonless configurations is illustrated in the middle panel of figure~\ref{fig:metric0}. 
Then, following the approach used in the previous sections, we consider the gravitational collapse of a thin shell to form the black hole described by \eqref{metric5d}.
The shell equation is given by
\begin{equation}
    \dot{R}^2-\frac{\pi \rho_{0} R^{2}\left[6C_{1}^{2}\rho_{0}^{2/3}R^{4}+12C_{1}\rho_{0}^{1/3}R^{2}+\pi\rho_{0}R^{2}+6\right]}
{9\left(1+C_{1}\rho_{0}^{1/3}R^{2}\right)^{4}}=0.
\end{equation}

 The effective potential near the origin can be expanded as

\begin{equation}
V(R)=-\frac{2\pi\rho_{0}}{3}R^{2}+\left(\frac{2\sqrt{2}\,\pi^{2}\rho_{0}^{3/2}}{3\sqrt{M}}-\frac{\pi^{2}\rho_{0}^{2}}{9}\right)R^{4}+\mathcal{O}(R^{6}).
\end{equation}
Thus, the shell radius behaves as
\begin{equation}
R(\tau)\sim R_{0}\exp\left[-\sqrt{\frac{2\pi\rho_{0}}{3}}(\tau-\tau_{0})
\right].
\end{equation}
This shows that the shell approaches the origin $R=0$ in the infinite proper time.

\begin{figure}
\centering
\includegraphics[width=0.45\columnwidth]{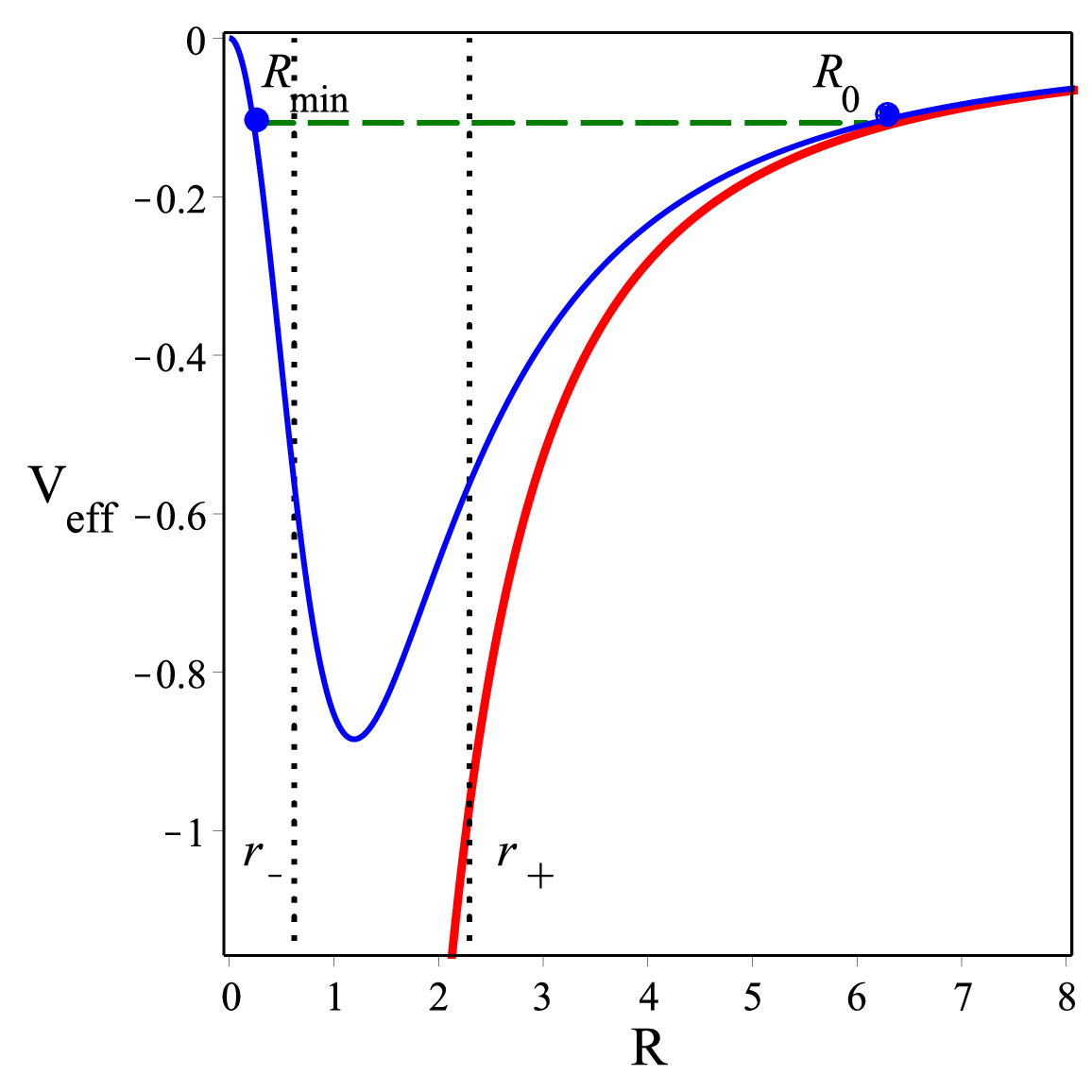}
\includegraphics[width=0.45\columnwidth]{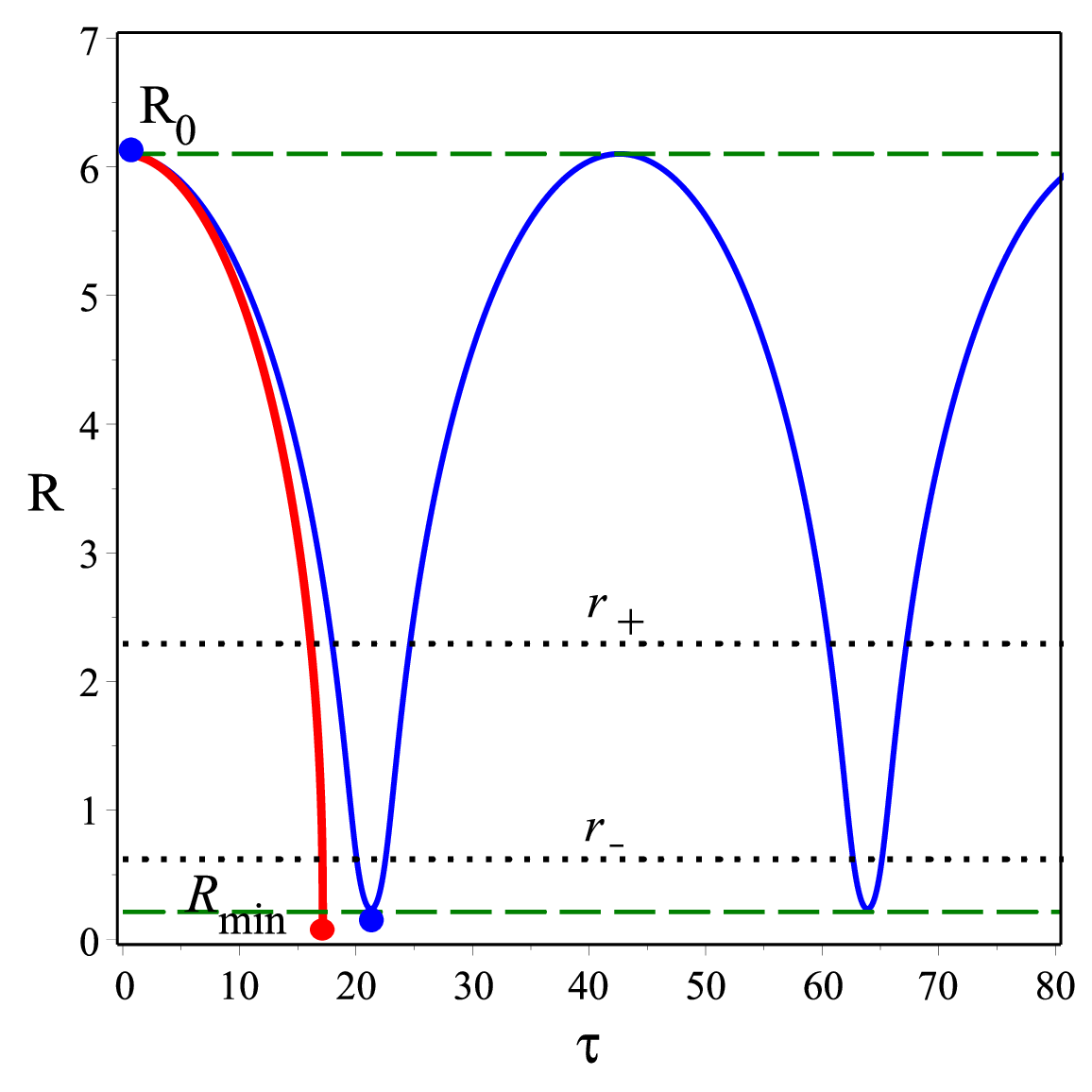}
\caption{Effective potential $V_{\rm eff}$ governing the thin-shell motion (left panel) and the shell radius $R$ as a function of proper time $\tau$ (right panel). The blue curves correspond to the regular black hole, while the red solid curves represent the Schwarzschild black hole. For the regular black hole, the shell collapses from $R_{0}=6.1$, crosses the horizons, reaches a finite minimum radius $R_{\min}=0.21$, and then bounces back. The shell subsequently expands into a new asymptotic region, and the collapse--bounce process repeats whenever the shell returns to $R_{0}$. In contrast, the Schwarzschild shell continues to collapse toward $R=0$, where the singularity is formed. The parameters are fixed as $D=5$, $\Lambda=0$, $\omega_{1}=1$, $n=4/3, M=10,\rho_{0}=1$ with $\zeta=-2/3$.
}
\label{fig:dynamic5d}
\end{figure}

Figure \ref{fig:dynamic5d} illustrates the collapse dynamics of a black hole similar to those in the previous sections. For the five-dimensional regular black hole, the shell crosses the horizons, reaches a finite minimum radius, and then bounces back, avoiding the formation of a singularity. The shell subsequently expands into a new asymptotic region, and the collapse--bounce process can repeat. In contrast, for the Schwarzschild black hole, the shell continues to collapse toward $R=0$, leading to a singularity. This comparison further highlights the regular nature of the black hole solution.

\subsection{The case $D=6$}\label{subsecD6}
\begin{figure}
\centering
\includegraphics[width=0.45\columnwidth]{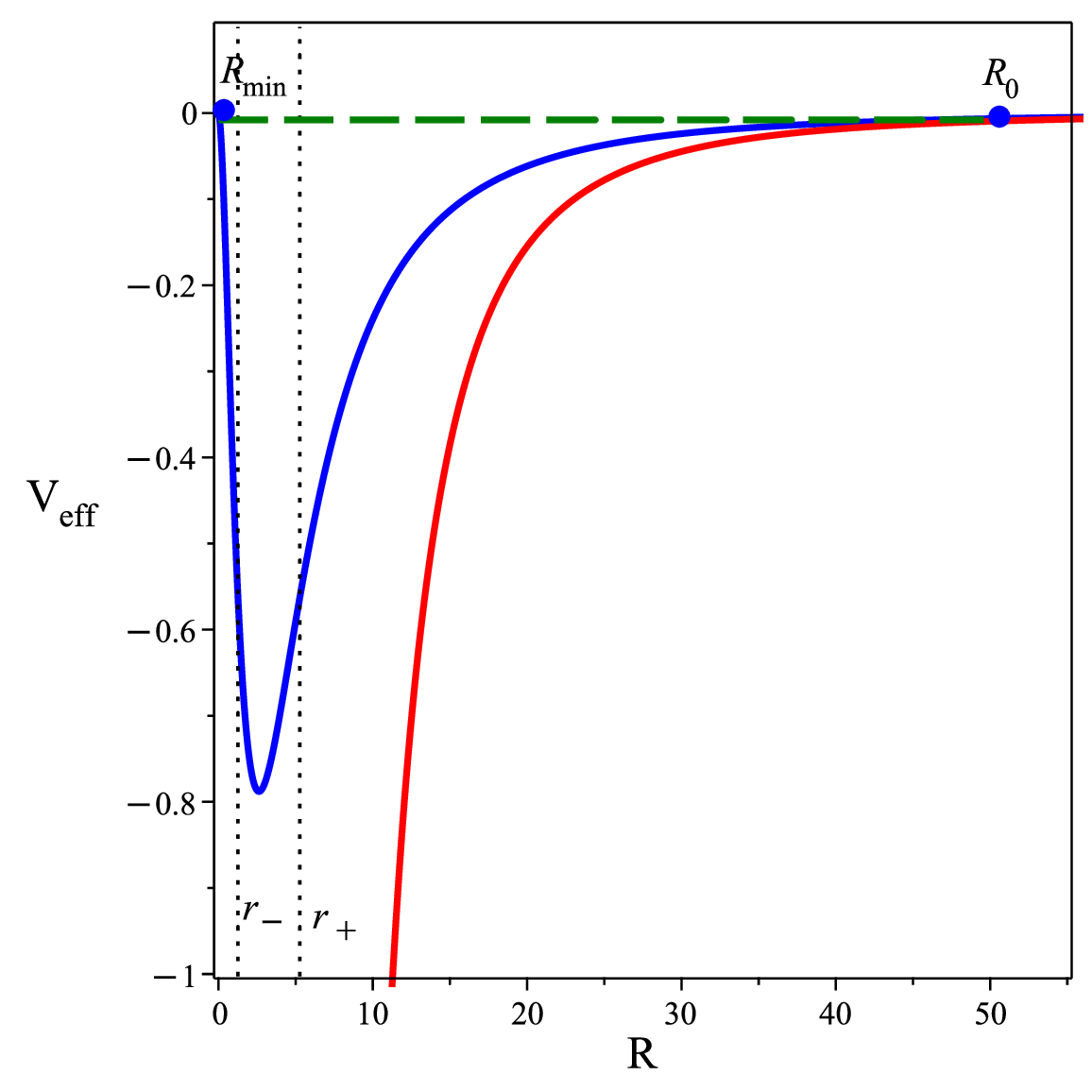}
\includegraphics[width=0.45\columnwidth]{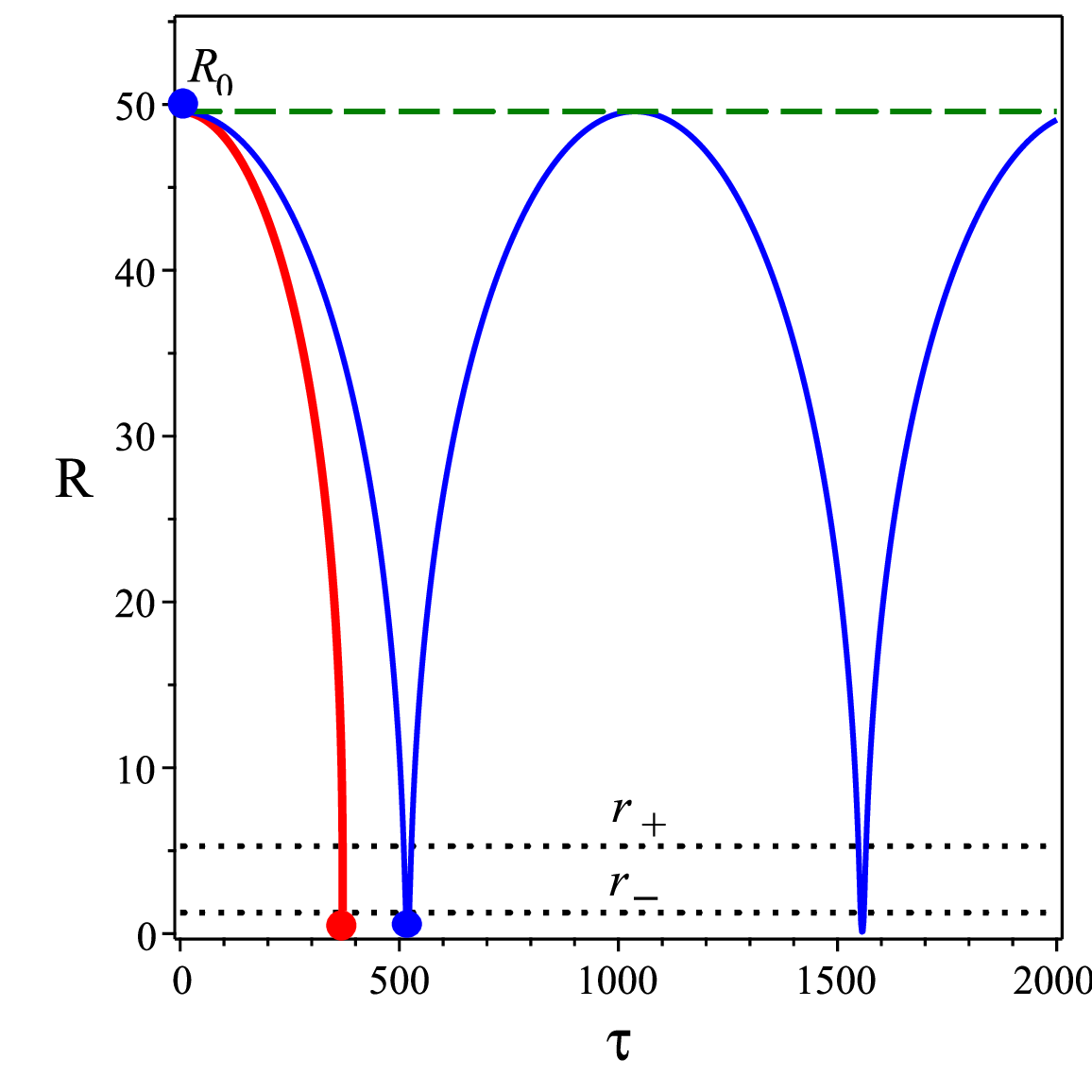}
\caption{Effective potential $V_{\rm eff}$ governing the thin-shell motion (left panel) and the shell radius $R$ as a function of proper time $\tau$ (right panel). The blue curves correspond to the regular black hole, while the red solid curves represent the Schwarzschild black hole. For the regular black hole, the shell collapses from $R_{0}=49.6$, crosses the horizons, reaches a finite minimum radius $R_{\min}=0.02$, and then bounces back. The shell subsequently expands into a new asymptotic region, and the collapse--bounce process repeats whenever the shell returns to $R_{0}$. In contrast, the Schwarzschild shell continues to collapse toward $R=0$, where the singularity is formed. The parameters are fixed as $D=6$, $\Lambda=0$, $\omega_{1}=1/2$, $n=7/6, M=5000,\rho_{0}=1$ with $\zeta=-3/4$.}
\label{fig:dynamic6d}
\end{figure}

For the six-dimensional case with \(D=6\), \(n=\tfrac{7}{6}\), and
\(\omega_{1}=\tfrac{1}{2}\), the metric function can be expressed as
\begin{equation}\label{eqmetric6d}
f(r)=1-\frac{\Lambda r^{2}}{10}
-\frac{4\pi\rho_{0}r^{2}}
{5\left(1+C_{1}r\,\rho_{0}^{1/6}\right)^{5}},
\end{equation}
here
\begin{equation}\label{eqC16d}
C_{1}
=
\left(\frac{8\pi^{2}\rho_{0}^{1/6}}{15M}\right)^{1/5}.
\end{equation}
The behavior of the metric function \eqref{eqmetric6d} is illustrated in Figure \eqref{fig:metric0} for different values of the parameters. In each panel, increasing the mass parameter shifts the metric function upward, leading to the formation of event horizons. The lowest-mass curve corresponds to a horizonless, regular spacetime, while the intermediate curve represents the extremal black hole with a degenerate horizon. For larger masses, the metric function possesses two distinct zeros, corresponding to the inner (Cauchy) and outer (event) horizons. These features show that the regular black hole solutions exhibit the same qualitative horizon structure in different spacetime dimensions, although the critical mass required for horizon formation increases significantly with the dimension.\\
Next, similar to four and five dimensional black holes in the previous sections, we will study the collapse of black hole. 
The shell equation is given by

\begin{equation}
\dot{R}^2-\frac{2\pi\rho_0 R^2}{5\left(1+1.393990117\left(\frac{\rho_0}{M}\right)^{1/5}R\right)^5}-\frac{\pi^2\rho_0^2R^4}{50\left(1+1.393990117\left(\frac{\rho_0}{M}\right)^{1/5}R\right)^{10}}=0.
\end{equation}
The near origin behavior of the effective potential is given by
\begin{align}
V_{\rm eff}(R)={}&-\frac{2\pi\rho_0}{5}R^2+\frac{2.787980234\,\pi\rho_0^{6/5}}{M^{1/5}}R^3+\mathcal{O}(R^4).
\end{align}
Hence, the near-origin evolution of shell radius is
\begin{equation}
R(\tau)\sim R_0
\exp\left[-\sqrt{\frac{2\pi\rho_{0}}{5}}(\tau-\tau_0)\right].
\end{equation}
This shows that, in the near-origin approximation, the shell approaches the origin asymptotically rather than reaching $R=0$ in finite proper time. The behavior of collapse of a thin shell has been shown in figure \ref{fig:dynamic6d}. The shell starts at $R=R_{0}$ and contracts toward smaller radii. It reaches a finite minimum radius $R_{\min}=0.02$, where $\dot{R}=0$, and then reverses its motion and expands again. Thus, the shell does not reach the origin; instead, it undergoes a bounce at a finite radius. The subsequent evolution repeats when the shell returns to $R_{0}$.

The difference between the regular black hole metric and the corresponding Schwarzschild--Tangherlini metric becomes larger in the small and intermediate radius regions as the spacetime dimension increases, since the Schwarzschild term behaves as $r^{-(D-3)}$, while the regular solution remains finite at the origin.

In this section, we have shown that these regular black holes are dynamically compatible with gravitational collapse under the thin-shell model. In the next section, we study other properties of these black holes, focusing on their thermodynamics and possible phase transitions.

\section{Thermodynamics of regular black hole}\label{thermosection}

In this section, we study the thermodynamic behavior of black holes in various spacetime dimensions \cite{thermorl}. We verify the first law of black hole thermodynamics and the Smarr relation.

Starting from the \(tt\)-component of the Einstein equations, and using the asymptotic behavior of the metric together with the horizon condition \(f(r_h)=0\), one can rewrite the total mass as the sum of the mass inside and outside the horizon, as follows \cite{Hennigar:2025yqm}
\begin{equation}
M=\frac{(D-2)\Omega_{D-2}}{16\pi}\left(r_h^{D-3}+\frac{r_h^{D-1}}{\ell^2}\right)+\Omega_{D-2}\int_{r_h}^{\infty}r^{D-2}\rho(r)\,dr.
\end{equation}
By using the energy density \eqref{eqqrho} and the incomplete Beta function\footnote{The incomplete Beta function is defined by
\begin{equation}
B_z(a,b)=\int_0^zt^{a-1}(1-t)^{b-1}\,dt.
\end{equation}}, one can explicitly write the total mass as follows
\begin{equation}
\begin{aligned}
M=\frac{(D-2)\Omega_{D-2}}{16\pi}\left(r_h^{D-3}+\frac{r_h^{D-1}}{\ell^2}\right)+\frac{\Omega_{D-2}\rho_0\left(C_1\rho_0^{\,n-1}\right)^{-\frac{D-1}{\alpha}}}{\alpha}B_{z}\left(\frac{1}{n-1}-\frac{D-1}{\alpha},\,\frac{D-1}{\alpha}\right),
\end{aligned}
\end{equation}
where we have used
\begin{equation}
z=\frac{1}
{1+C_1\rho_0^{\,n-1}r_h^{\alpha}},\qquad\Lambda=-\frac{(D-1)(D-2)}{2\ell^2}.
\end{equation}
Thus, the thermodynamic pressure and volume are
\begin{equation}
\mathcal{P}=-\frac{\Lambda}{8\pi}=\frac{(D-1)(D-2)}{16\pi \ell^2},\qquad V=\frac{\Omega_{D-2}}{D-1}r_h^{D-1}.
\end{equation}
Since the gravitational sector is described by Einstein gravity and the matter is
minimally coupled, the black hole entropy satisfies the usual area law,
\begin{equation}
S=\frac{\Omega_{D-2}}{4}r_h^{D-2}.
\end{equation}

The Hawking temperature, using the thermodynamic pressure and using the derivative identity of the hypergeometric function, is given by\footnote{The derivative of the hypergeometric function is
\begin{equation}
\frac{d}{dz}\,{}_2F_1(a,b;1+b;-z)=-\frac{ab}{1+b}\,{}_2F_1(a+1,b+1;2+b;-z).
\end{equation}}

\begin{align}
T=&\dfrac{f^{\prime}(r_{h})}{4\pi}=\frac{8\mathcal{P}r_h}{(D-1)(D-2)}
-\frac{8\rho_0 r_h}{(D-1)(D-2)}\,{}_2F_1\left(\frac{1}{n-1},\frac{D-1}{\alpha};
1+\frac{D-1}{\alpha};-C_1\rho_0^{\,n-1}r_h^{\alpha}\right)
\nonumber\\
&+\frac{4C_1\rho_0^{\,n}r_h^{\alpha+1}}{(D-2)(n-1)\left(1+\frac{D-1}{\alpha}\right)}
{}_2F_1\left(\frac{n}{n-1},1+\frac{D-1}{\alpha};2+\frac{D-1}{\alpha};
-C_1\rho_0^{\,n-1}r_h^{\alpha}\right).
\end{align}
In the extended phase space, the first law of black hole thermodynamics takes the form
\begin{equation}
 \mathcal{A}\delta M=T\delta S+V\delta \mathcal{P}+\Psi_{\rho_{0}}\delta \rho_{0},
\end{equation}
and the Smarr formula is given by
\begin{equation}
    {(D-3)M=(D-2)TS-2\mathcal{P}V+\Delta},
\end{equation}
where $\mathcal{A}$, $\Psi_{\rho_{0}}$ and $\Delta$ are given by \eqref{firstlaw}, \eqref{eq:pressureintegral} and \eqref{eqdelta} in appendix \ref{appsmarr}.

\subsection{Stability and phase transition}

Here, we investigate possible phase transitions and the thermodynamic stability of the black holes presented in subsections \ref{subsecD4} to \ref{subsecD6} using the free energy and the heat capacity in the canonical ensemble in four, five, and six spacetime dimensions.

\subsection{The case $D=4$, \(n=\frac{3}{2}\), \(\omega_{1}=2\)}

The thermodynamical quantities corresponding to this case are given by
\begin{align}
&T=-\frac{64\pi^2\mathcal{P}^2r_h^4-64\pi^2\mathcal{P}\rho_0r_h^4
+48\pi \mathcal{P} r_h^2-8\pi\rho_0 r_h^2+9}{32\pi^2\rho_0 r_h^3},\quad S=\pi r_{h}^2,\quad V=\dfrac{4}{3}\pi r_{h}^3,\nonumber\\
&\mathcal{P}=-\dfrac{\Lambda}{8\pi},\quad M=\frac{4\pi\rho_{0}r_{h}^{3}}{3}\,\frac{3+8\pi \mathcal{P} r_{h}^{2}}
{-3+8\pi (\rho_{0}-\mathcal{P})r_{h}^{2}},\quad\mathcal{A}=\frac{\left[3+8\pi(\mathcal{P}-\rho_0)r_h^2\right]^2}
{64\pi^2\rho_0^2 r_h^4},\nonumber\\
&\Delta=\frac{\left(9+24\pi \mathcal{P} r_h^2-32\pi\rho_0 r_h^2\right)
\left(3+8\pi \mathcal{P} r_h^2\right)^2}
{48\pi\rho_0 r_h
\left[3+8\pi(\mathcal{P}-\rho_0)r_h^2\right]},\qquad \Psi_{\rho_0}=-\frac{\left(3+8\pi \mathcal{P} r_h^2\right)^2}
{48\pi\rho_0^{2}r_h}.
\end{align}
It is straightforward to show that the first law and the Smarr formula take the form
\begin{align}
   \mathcal{A}\delta M &= T\delta S+V\delta \mathcal{P}+\Psi_{\rho_{0}}\delta\rho_{0},\\
    M &= 2TS-2\mathcal{P}V+\Delta.
\end{align}
We analyze the critical behavior of the system in the standard manner by requiring that the equation of state satisfy the critical (inflection) point conditions:
\begin{equation}
    \left(\dfrac{\partial T}{\partial r_{h}}\right)_{\mathcal{P},\rho_{0}}=\left(\dfrac{\partial^{2}T}{\partial r_{h}^2}\right)_{\mathcal{P},\rho_{0}}=0.
\end{equation}
The black hole has a single critical point with the following critical parameters:
\begin{equation}
    \mathcal{P}_{c} = 0.0084\rho_{0},\quad r^{c}_{h} = \dfrac{1.5043}{\sqrt{\rho_{0}}},\quad T^{c}=0.0669\sqrt{\rho_{0}}.
\end{equation}
The dimensionless ratio of these critical parameters is given by
\begin{equation}
    \dfrac{\mathcal{P}_{c} r^{c}_{h}}{T_{c}}=0.1891.
\end{equation}
Since this ratio is independent of the overall scale of the critical parameters, it provides a useful characterization of the critical behavior of the system. Its numerical value is close to, but differs slightly from, the van der Waals value $0.1875$.
This small deviation indicates that, although the black hole undergoes a van der Waals-like phase transition, its critical behavior is modified by the underlying matter distribution and the corresponding spacetime geometry.

Using the first law of thermodynamics, we define the corrected internal energy by integrating
\begin{equation}\label{eqE64generic}
    E=\int \mathcal{A}\,dM
    =\int \mathcal{A}\,\left(\frac{dM}{dr_{h}}\right)_{\mathcal{P},\rho_{0}}\,dr_{h},
\end{equation}
which yields
\begin{equation}
    E=\frac{r_{h}}{2}+\frac{9}{16\pi\rho_{0}r_{h}}-\frac{4\pi}{3\rho_{0}}\,\mathcal{P}^{2}r_{h}^{3}+\frac{4\pi}{3}\,\mathcal{P} r_{h}^{3}-\frac{3\mathcal{P}r_{h}}{\rho_{0}}.
\end{equation}
The corresponding free energy is then obtained from
\begin{equation}
    F=E-TS,
\end{equation}
which can be written as
\begin{equation}
    F=
    \frac{r_{h}}{4}
    +\frac{27}{32\pi\rho_{0}r_{h}}
    +\frac{2\pi\mathcal{P}^{2}r_{h}^{3}}{3\rho_{0}}
    -\frac{2\pi\mathcal{P}r_{h}^{3}}{3}
    -\frac{3\mathcal{P}r_{h}}{2\rho_{0}}.
\end{equation}

\begin{figure}
\centering
\includegraphics[width=0.32\columnwidth]{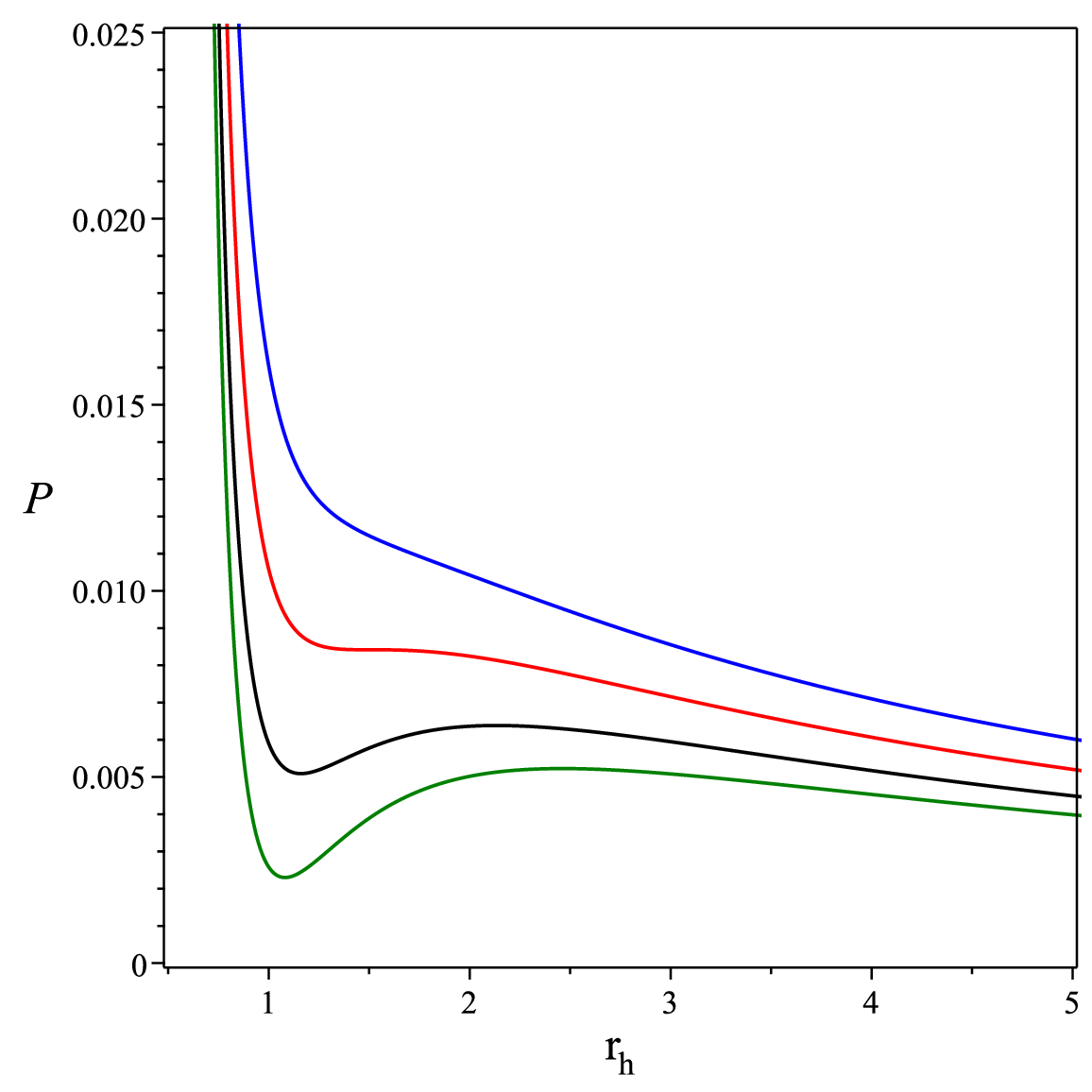}
\includegraphics[width=0.32\columnwidth]{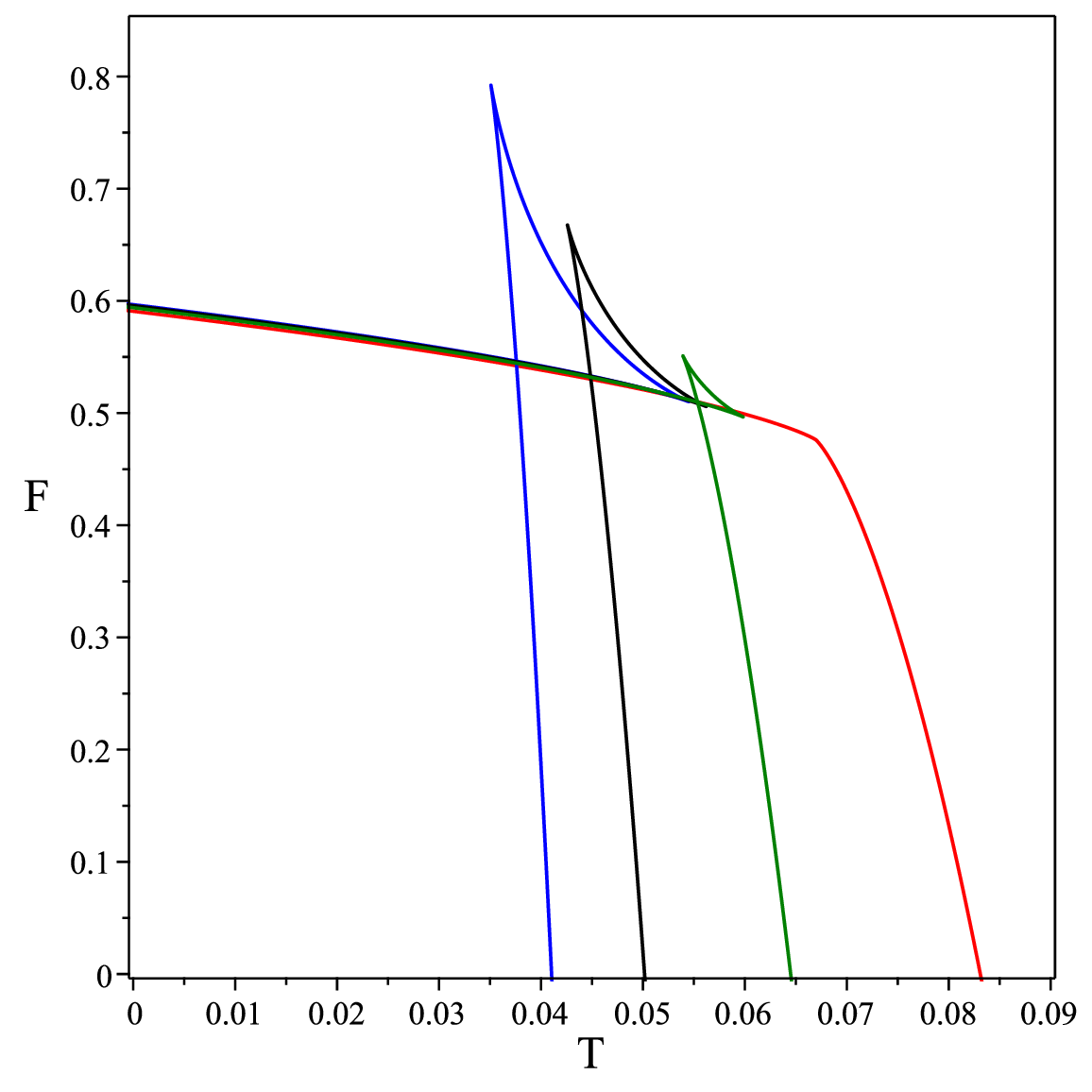}
\includegraphics[width=0.32\columnwidth]{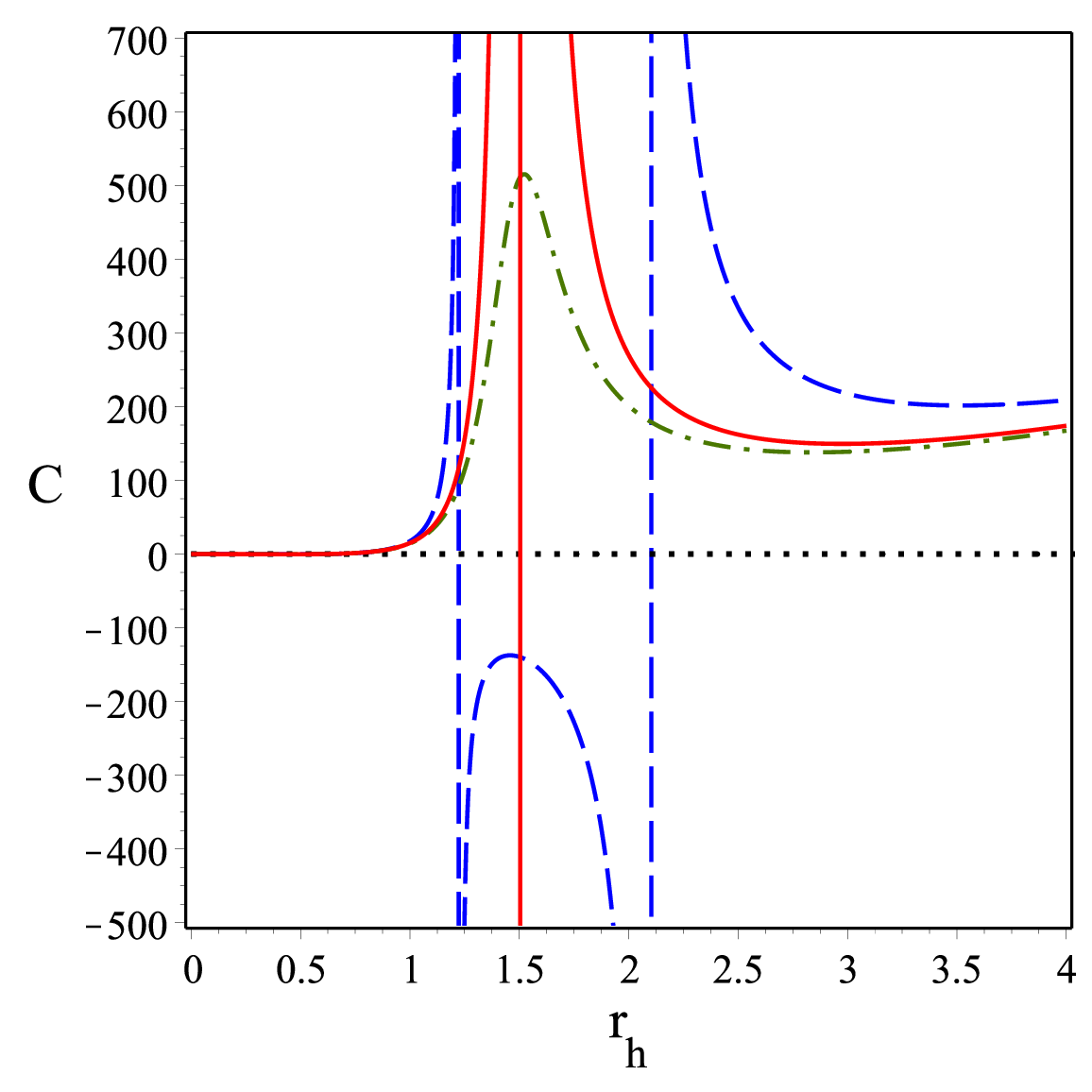}
\caption{Left: The black hole pressure \(\mathcal{P}\) as a function of the horizon radius \(r_h\) for \(D=4\) and \(T={0.055},\;0.06,\;{0.0669},\;{0.075}\) (from bottom to top). Middle: Free energy \(F\) as a function of temperature \(T\) for pressures \(\mathcal{P}={0.002}\), \(0.003\), \({0.005}\), and \({0.0084}\). Right: Heat capacity as a function of the horizon radius $r_{h}$ and $\mathcal{P}={0.0065}$, ${0.0084}$, and ${0.009}$. We have set $\rho_{0}=1$.}
\label{fig:prhplot}
\end{figure}

The phase structure of the black hole is analyzed through the equation of state
and free energy. In the left panel of figure.~\ref{fig:prhplot}, the
pressure $\mathcal{P}$ is plotted as a function of the horizon radius $r_h$ for
different temperatures. For $T<T_c$, the isotherms exhibit the characteristic
oscillatory behavior of a van der Waals system, indicating a first-order phase
transition between small and large black holes. At critical temperature
$T_c=0.0669$, the oscillation disappears and the isotherm develops an
inflection point.
For $T>T_c$, the pressure becomes monotonic and no phase transition occurs.
The middle panel shows the free energy as a function of temperature. The
swallowtail structure for $\mathcal{P}<\mathcal{P}_c$ confirms the existence of
a first-order small black hole/large black hole phase transition. At critical pressure $\mathcal{P}_c=0.0084$, the swallowtail disappears,
signaling the second-order critical point. These results demonstrate that the
black hole system exhibits a van der Waals-like phase transition.

The right panel of figure~\ref{fig:prhplot} shows the behavior of the heat capacity $C_{\mathcal{P},\rho_{0}}$ as a function of the horizon radius $r_h$ for $\mathcal{P}=0.0065$, $0.0084$, and $0.009$. The heat capacity at constant pressure is defined as
\begin{equation}
C_{\mathcal{P},\rho_{0}}= T\left(\frac{\partial S}{\partial T}\right)_{\mathcal{P},\rho_{0}}
\end{equation}
where $T$ is the Hawking temperature and $S$ is the black hole entropy. The heat capacity provides information on the local thermodynamic stability of the black hole. In particular, $C_{\mathcal{P},\rho_{0}}>0$ indicates a locally stable black hole, while $C_{\mathcal{P},\rho_{0}}<0$ indicates a locally unstable one.

As shown in the figure, the heat capacity can diverge at certain values of the horizon radius. These divergences occur when
\begin{equation}
\left(\frac{\partial T}{\partial r_h}\right)_{\mathcal{P},\rho_{0}}=0.
\end{equation}
Therefore, the divergent points correspond to extrema of the Hawking temperature and can be associated with second order phase transitions.

For the blue dashed curve, the heat capacity has two divergent points. Between these two points, the heat capacity is negative, indicating an unstable intermediate black hole branch. On both sides of this region, the heat capacity is positive, corresponding to locally stable black hole configurations. Thus, the black hole shows a stable/unstable/stable behavior as $r_h$ increases.
The positions of the divergent points depend on the pressure. As the pressure is varied, these points move and can eventually merge at a critical pressure.

\subsection{The case $D=5$, $n=\frac{4}{3}$, $\omega_{1}=1$}
The thermodynamic quantities, expressed as a function of the radius of the horizon, the pressure, and the density of matter, are given by

\begin{align}
&T=\frac{2\mathcal{P} r_h}{3}
-\frac{2\rho_0 r_h}{3\left(1+C_1 r_h^2 \rho_0^{1/3}\right)^2}
+\frac{4C_1 \rho_0^{4/3} r_h^3}{3\left(1+C_1 r_h^2 \rho_0^{1/3}\right)^3},\qquad S=\frac{\pi^{2}r_{h}^{3}}{2},\qquad V=\frac{\pi^{2}r_{h}^{4}}{2},\nonumber\\
&\mathcal{P}=-\frac{\Lambda}{8\pi},\qquad 
\mathcal{A}=
\frac{C_{1}^{3}\rho_{0}r_{h}^{6}}
{\left(1+C_{1}\rho_{0}^{1/3}r_{h}^{2}\right)^{3}},\qquad \Delta=
\frac{\pi^{2}\rho_{0}^{1/3}}{C_{1}^{2}}
\frac{
1+3C_{1}\rho_{0}^{1/3}r_{h}^{2}
+4C_{1}^{2}\rho_{0}^{2/3}r_{h}^{4}
}
{\left(1+C_{1}\rho_{0}^{1/3}r_{h}^{2}\right)^{3}},
\end{align}

\begin{equation}
M=\frac{\pi^{2}r_{h}^{4}\left(4\pi \mathcal{P} r_{h}^{2}+3\right)
\left[
4\pi \mathcal{P}\rho_{0}r_{h}^{2}
+4\pi\rho_{0}^{2}r_{h}^{2}
+4\rho_{0}^{3/2}
\sqrt{\pi r_{h}^{2}\left(4\pi \mathcal{P} r_{h}^{2}+3\right)}
+3\rho_{0}
\right]}
{2\left(4\pi \mathcal{P} r_{h}^{2}-4\pi\rho_{0}r_{h}^{2}+3\right)^{2}},
\end{equation}

\begin{equation}
\Psi_{\rho_{0}}=-\frac{\sqrt{\pi}\,r_h\left(4\pi \mathcal{P} r_h^2+3\right)^{3/2}}
{16\,\rho_0^{3/2}},~~~~C_{1}=
-\frac{1}{\rho_{0}^{1/3}r_{h}^{2}}
+\frac{2\sqrt{\pi}\,\rho_{0}^{1/6}}
{r_{h}\sqrt{3+4\pi \mathcal{P} r_{h}^{2}}}.
\end{equation}

It is straightforward to verify that the above thermodynamic quantities satisfy the first law of black hole thermodynamics and the Smarr relation
\begin{align}
   \mathcal{A}\delta M=&T\delta S+V\delta \mathcal{P}+\Psi_{\rho_{0}}\delta\rho_{0},\\
    2M =&3TS-2\mathcal{P}V+\Delta.
\end{align}
The critical point is determined by the inflection conditions of the equation of state and is characterized by
\begin{equation}
\mathcal{P}_{c}=0.0047\,\rho_{0},\qquad
r_{h}^{c}=\frac{2.9408}{\sqrt{\rho_{0}}},\qquad
T_{c}=0.0498\,\sqrt{\rho_{0}}.
\end{equation}
The dimensionless ratio
\begin{equation}
    \frac{\mathcal{P}_{c} r_h^c}{T_c}=0.2787,
\end{equation}
describes the critical behavior of the five-dimensional regular black hole for
$n=\frac{4}{3}$ and $\omega_1=1$. This value is different from the four-dimensional
case, showing that the critical ratio depends on the spacetime dimension and the
matter source parameters. Thus, the critical ratio is not universal for the regular
black holes considered here.

\begin{figure}
\centering
\includegraphics[width=0.32\columnwidth]{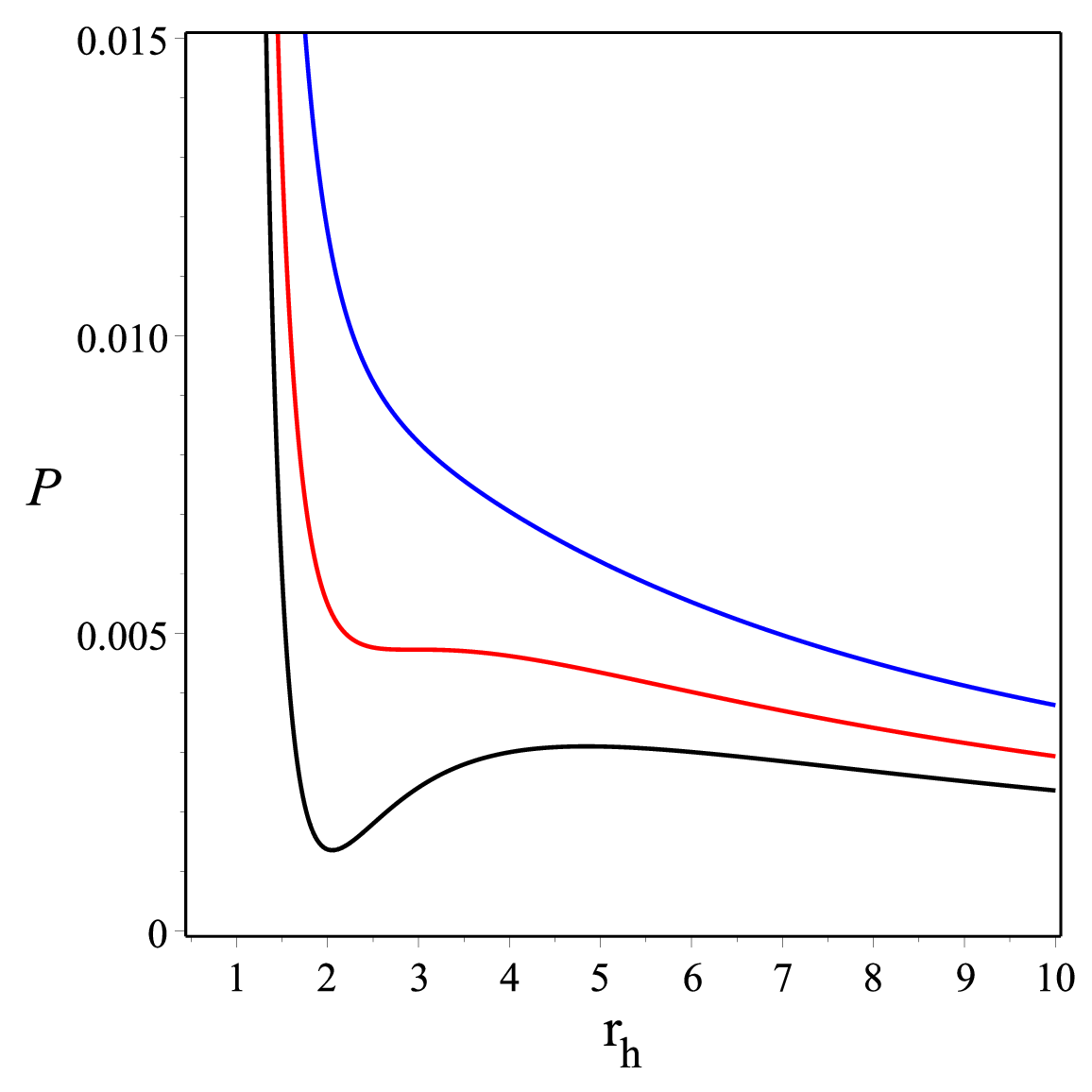}
\includegraphics[width=0.32\columnwidth]{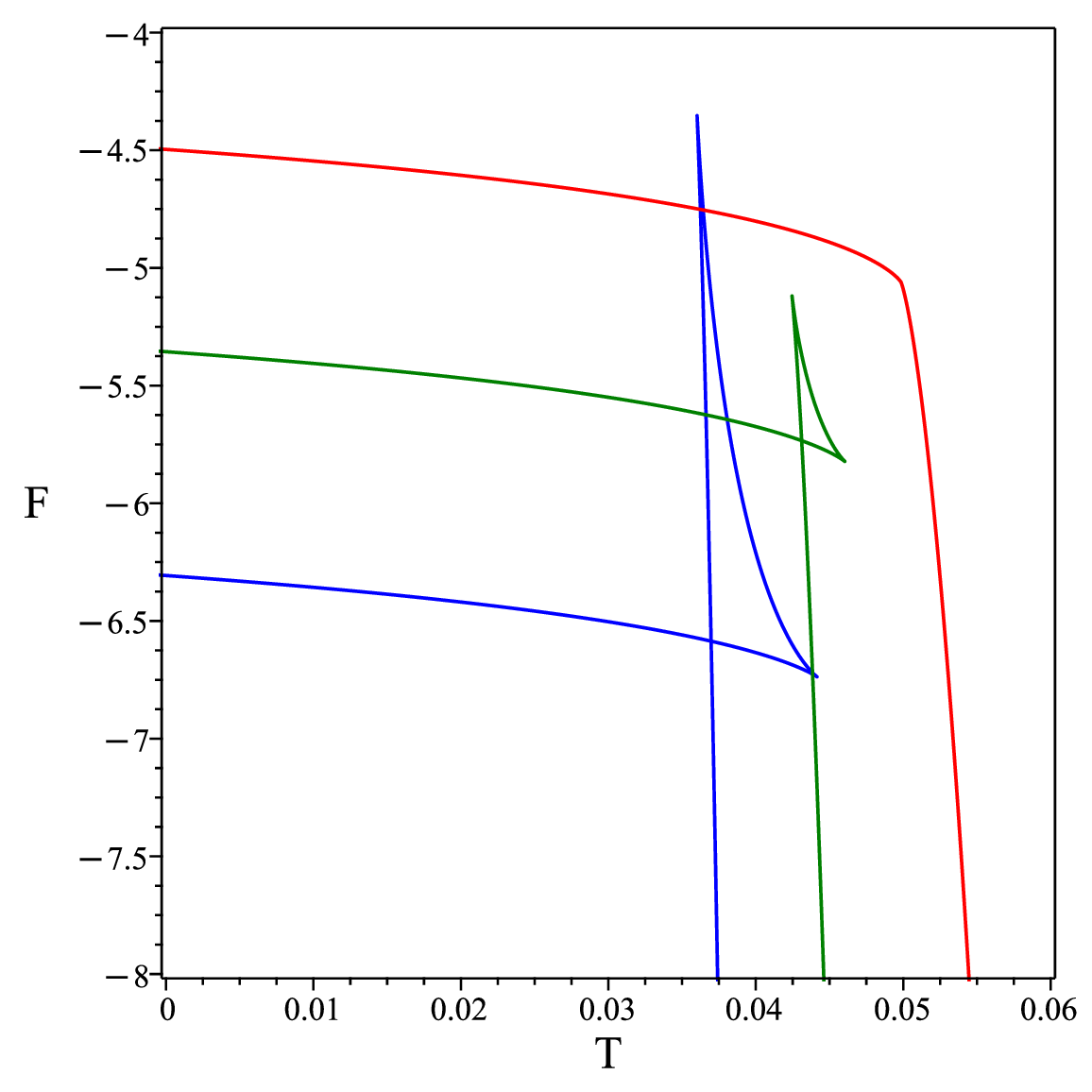}
\includegraphics[width=0.32\columnwidth]{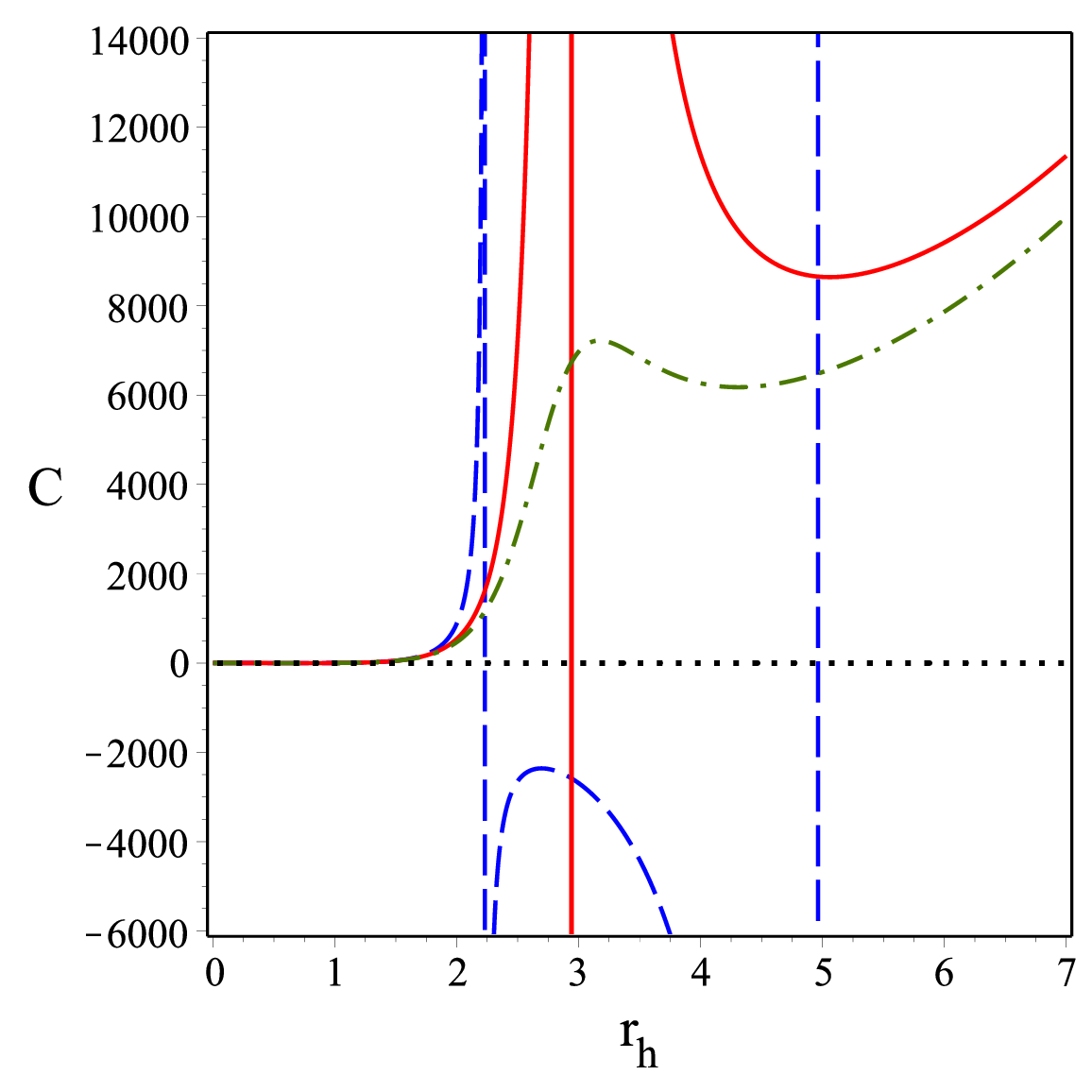}
\caption{Left: The black hole pressure \(\mathcal{P}\) as a function of the horizon radius \(r_h\) for \(D=5\) and \(T=0.043,\;{0.0498},\;{0.06}\) (from bottom to top). {Middle:} Free energy as a function of temperature \(T\) for pressures \(\mathcal{P}={0.002}\), \({0.003}\), and \({0.0047}\) (from bottom to top). Right: Heat capacity in terms of $r_{h}$ for $\mathcal{P}={0.003}$~(\text{dash line}), ${0.0047}~(\text{solid line})$, and ${0.0055}~(\text{dash-dot line})$. Here we have set $\rho_{0}=1$.}
\label{fig:prhplot5d}
\end{figure}

Figure~\ref{fig:prhplot5d} demonstrates the thermodynamic behavior of the five-dimensional regular black hole. In the left panel, the equation of state exhibits the characteristic oscillatory behavior for temperatures below the critical temperature, signaling the coexistence of small and large black hole phases and the onset of a first-order phase transition. As the temperature approaches the critical value, the oscillation shrinks to an inflection point, while for $T>T_c$ the isotherms become monotonic, indicating the disappearance of the phase transition.

The middle panel displays the free energy as a function of temperature for several fixed pressures. For \(\mathcal{P}<\mathcal{P}_c\), the characteristic swallowtail structure confirms the first-order transition between the small and large black hole phases. At critical pressure, the swallowtail disappears, signaling a second-order critical point. It is worth emphasizing that the free energy remains negative throughout the entire temperature range considered. Therefore, unlike the corresponding four-dimensional case, where the swallowtail may extend into the positive free energy region and allow a Hawking--Page transition, the five-dimensional black hole is always thermodynamically favored over thermal AdS. Consequently, the thermodynamic behavior is governed solely by the small/large black hole phase transition.

The right panel of the figure shows the heat capacity $C$ as a function of the horizon radius $r_h$. The divergences of $C$ indicate the presence of phase transitions, while its sign determines the local thermodynamic stability of the black holes. In particular, configurations with $C>0$ are thermodynamically stable, whereas those with $C<0$ are unstable.

\subsection{The case $D=6$, \(n=\tfrac{7}{6}\), \(\omega_{1}=\tfrac{1}{2}\)}
In this case, the thermodynamical quantities are given by
\begin{align}
&T=\frac{2\mathcal{P}r_{h}}{5}
-\frac{2\rho_{0}r_{h}}
{5\left(1+C_{1}r_{h}\rho_{0}^{1/6}\right)^{5}}
+\frac{C_{1}\rho_{0}^{7/6}r_{h}^{2}}
{\left(1+C_{1}r_{h}\rho_{0}^{1/6}\right)^{6}},\quad S=\frac{2\pi^{2}r_{h}^{4}}{3},\quad V=\frac{8\pi^{2}r_{h}^{5}}{15},\nonumber\\
&\mathcal{P}=-\frac{\Lambda}{8\pi},\quad\mathcal{A}
=\left(
\frac{C_{1}r_{h}\rho_{0}^{1/6}}
{1+C_{1}r_{h}\rho_{0}^{1/6}}
\right)^{6},\quad \Delta=\frac{4\pi^{2}\rho_{0}^{1/6}}{15C_{1}^{5}}
\left[3-\frac{C_{1}^{5}\rho_{0}^{5/6}r_{h}^{5}
\left(C_{1}\rho_{0}^{1/6}r_{h}+6\right)}
{\left(1+C_{1}\rho_{0}^{1/6}r_{h}\right)^{6}}
\right],\nonumber\\
& M=\frac{4\pi^{2}\rho_{0}r_{h}^{5}\left(10-\Lambda r_{h}^{2}\right)}{75\left[2\sqrt[5]{15\pi\rho_{0}r_{h}^{2}}-\left(10-\Lambda r_{h}^{2}\right)^{1/5}\right]^{5}
},\quad \Psi_{\rho_{0}}=-\frac{4\pi^2 C_1^6 r_h^{11}\rho_0
(10-\Lambda r_h^2)^{6/5}}
{75(1+C_1 r_h\rho_0^{1/6})^6\mathcal{D}^6},
\end{align}
here 
\begin{equation}
C_{1}=\frac{1}{r_h\rho_0^{1/6}}\left[\left(\frac{8\pi\rho_{0}r_{h}^{2}}{10-\Lambda r_{h}^{2}}\right)^{1/5}-1\right],\qquad \mathcal{D}=2(15\pi\rho_0 r_h^2)^{1/5}
-(10-\Lambda r_h^2)^{1/5}.
\end{equation}
It is straightforward to verify that the first law of thermodynamics and the Smarr relation take the following forms:
\begin{align}
\mathcal{A}\delta M &= T\delta S+V\delta \mathcal{P}
+\Psi_{\rho_{0}}\delta\rho_{0},\\
3M &= 4TS-2\mathcal{P}V+\Delta .
\end{align}
The critical thermodynamic quantities are found to be
\begin{equation}
\mathcal{P}_{c}=7.9867\times10^{-4}\rho_{0},
\qquad
r_{h}^{c}=\frac{8.0899}{\sqrt{\rho_{0}}},
\qquad
T_{c}=0.01542\sqrt{\rho_{0}} .
\end{equation}
The corresponding ratio of the critical quantities is given by
\begin{equation}
    \frac{\mathcal{P}_{c} r_{h}^{c}}{T_{c}}
    =0.4191.
\end{equation}

The critical ratio $\mathcal{P}_{c}r_h^c/T_c$ for different spacetime dimensions and model parameters is summarized in Table~\ref{tab:critical_ratio_dimension}. The ratio increases from $0.189$ in four dimensions to $0.2787$ in five dimensions and $0.4191$ in six dimensions. However, since the values of $n$ and $\omega_1$ are also varied with the spacetime dimension, this increase reflects the combined effects of the spacetime dimensionality and the properties of the matter source. Therefore, the critical ratio is not a universal quantity in the present class of regular black holes.

\begin{table}[h]
\centering
\begin{tabular}{|c|c|c|c|}
\hline
 $D$ 
&$n$ &$\omega_{1}$ & $\frac{\mathcal{P}_{c} r_h^c}{T_c}$ \\[3pt] 
\hline
4 &$\frac{3}{2}$ &2 & 0.1891 \\[6pt]

5 &$\frac{4}{3}$ & 1& 0.2787 \\[6pt]

6 &$\frac{7}{6}$ & $\frac{1}{2}$ & 0.4191 \\[6pt]
\hline
\end{tabular}
\caption{The critical ratio $\mathcal{P}_{c} r_h^c/T_c$ for different values of the model parameters $n$, $\omega_1$, and spacetime dimension $D$. The ratio depends on all three parameters, indicating that the thermodynamic phase structure of the regular black holes is influenced not only by the spacetime dimension but also by the properties of the matter source characterized by $n$ and $\omega_1$.
}
\label{tab:critical_ratio_dimension}
\end{table}

The corresponding free energy, derived from \eqref{eqE64generic}, is given by
\begin{equation}
\begin{aligned}
F={}&\frac{8\pi^{2}\mathcal{P}}{15}r_{h}^{5}
+\frac{2\pi}{3}r_{h}^{3}-\frac{2^{12/5}\pi^{4/5}}{3}
\left(\frac{10}{\rho_{0}}\right)^{1/5}\Bigg[\frac{5\pi\mathcal{P}}{23}r_{h}^{23/5}\,{}_2F_1\!\left(-\frac15,\frac{23}{10};\frac{33}{10};-\frac{4\pi\mathcal{P}}5r_{h}^{2}
\right)\\&\qquad\qquad+\frac{25}{52}r_{h}^{13/5}\,{}_2F_1\!\left(-\frac15,\frac{13}{10};
\frac{23}{10};-\frac{4\pi\mathcal{P}}5r_{h}^{2}\right)\Bigg].
\end{aligned}
\end{equation}

\begin{figure}
\centering
\includegraphics[width=0.32\columnwidth]{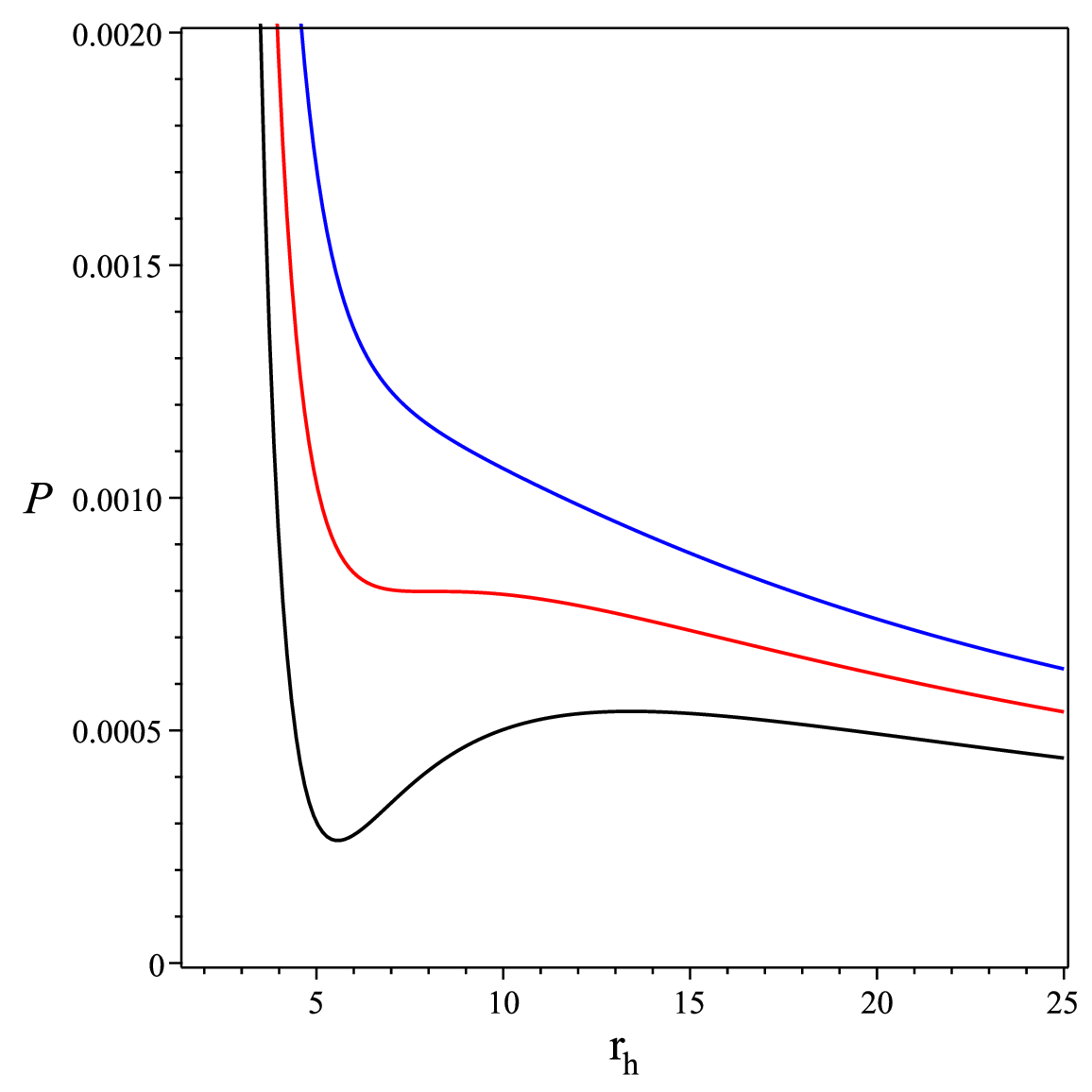}
\includegraphics[width=0.32\columnwidth]{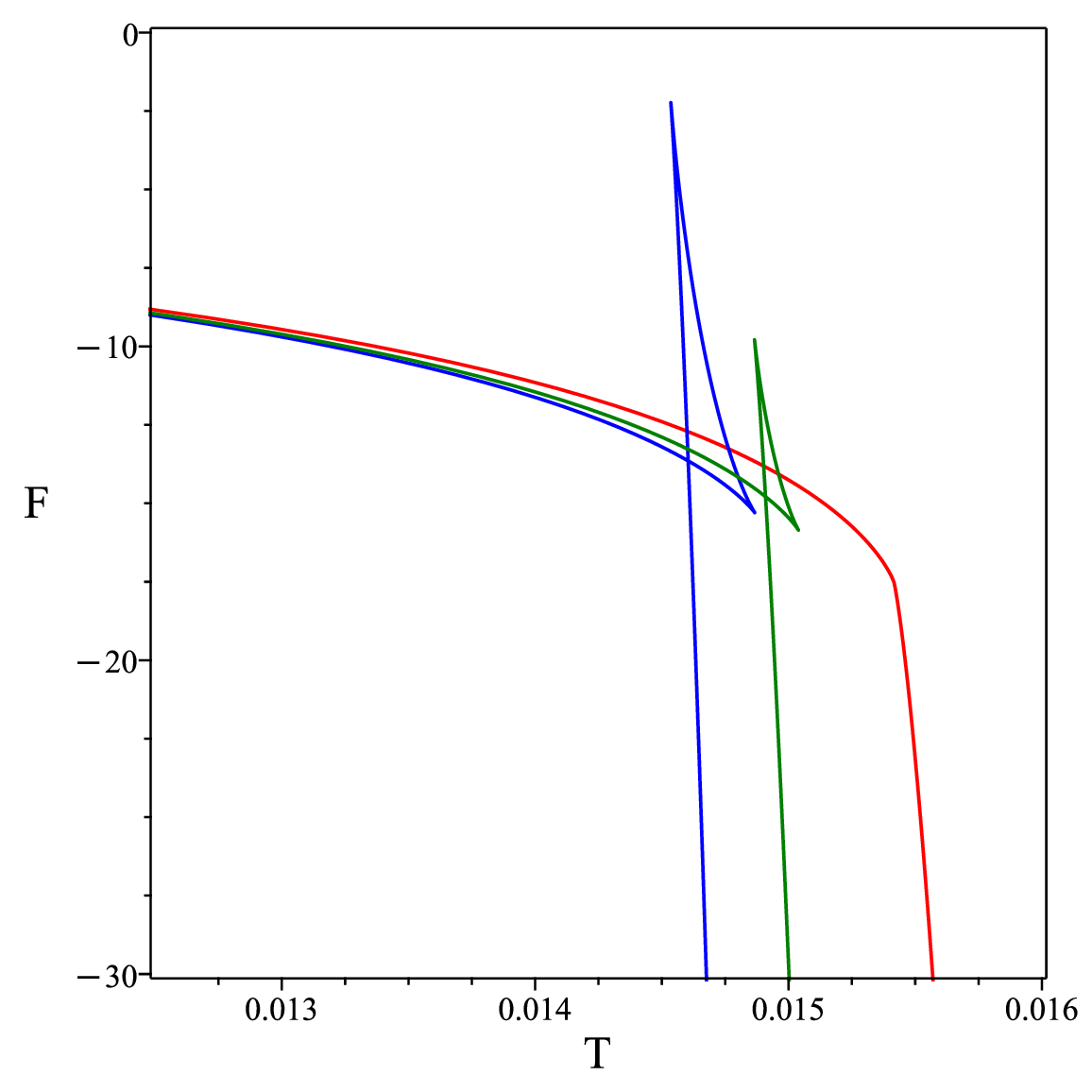}
\includegraphics[width=0.32\columnwidth]{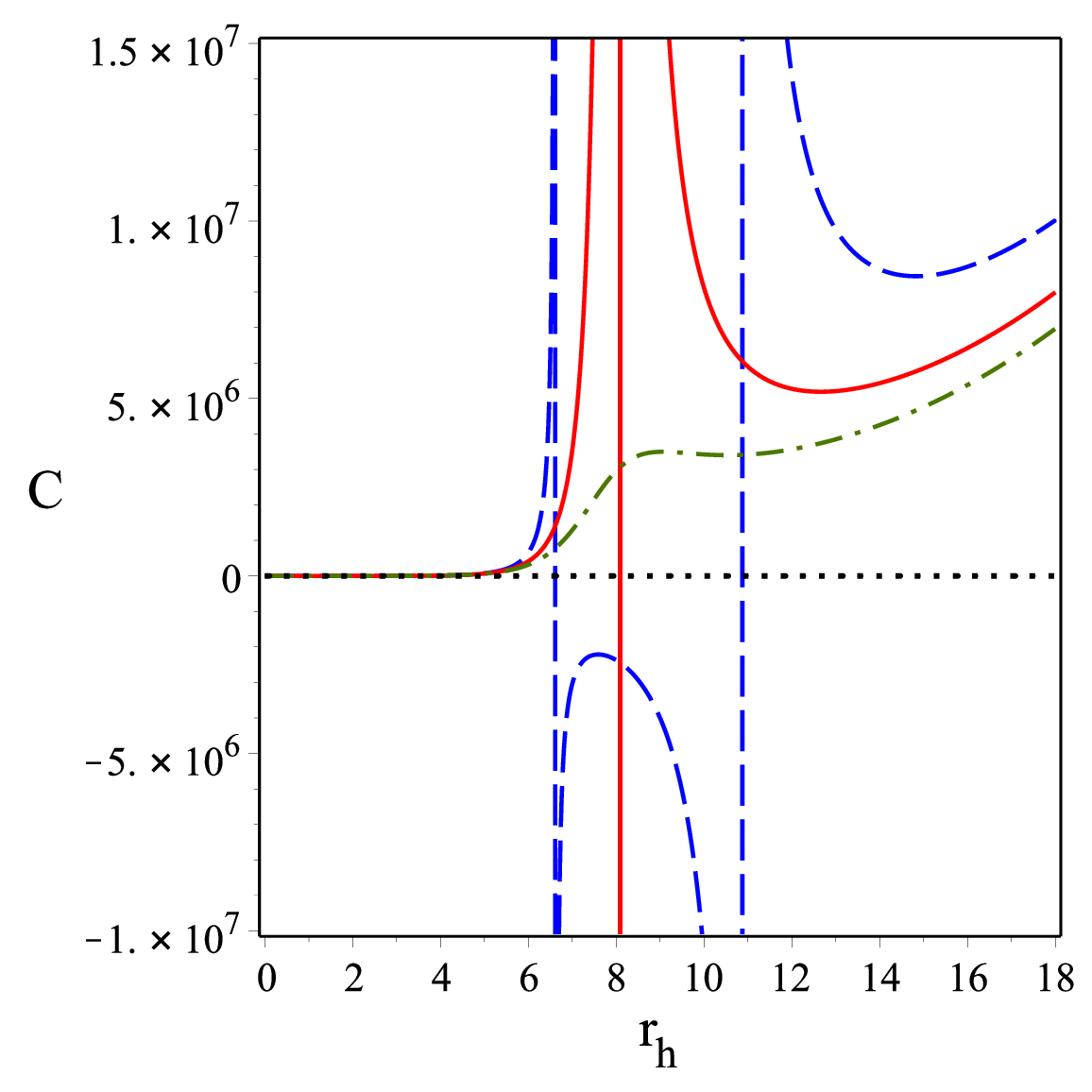}
\caption{Left: The black hole pressure \(\mathcal{P}\) as a function of the horizon radius \(r_h\) for \(D=6\) and \(T=0.0137,\;{0.01542},\;{0.017}\) (from bottom to top). {Middle:} Free energy \(F\) as a function of temperature \(T\) for pressures \(\mathcal{P}={0.00065}\), \({0.0007}\),  and \({0.00079867}\). Right: Heat capacity in terms of $r_{h}$ for $\mathcal{P}={0.00068}~(\text{dash line}),~{0.00079867}~(\text{solid line}),~{0.0009}~ (\text{dash-dot line})$. We have set $\rho_{0}=1$.}
\label{fig:prhplot6d}
\end{figure}

In figure~\ref{fig:prhplot6d}, we show the free energy as a function of temperature. Similar to the four- and five-dimensional cases, the black hole undergoes a small/large black hole phase transition. However, as in the five-dimensional case, the free energy remains negative for all temperatures, indicating that the black hole is thermodynamically favored throughout the entire temperature range. This behavior, reflected in both the left and right panels, resembles that observed in the four and five-dimensional cases.

\section{Conclusion}\label{secconc}
In this work, we have constructed a general class of static, spherically symmetric regular black hole solutions in arbitrary spacetime dimensions. The black holes are supported by an anisotropic fluid with vacuum-like radial pressure and a generalized polytropic equation of state for the tangential pressure. We obtain asymptotically anti-de Sitter solutions and identify the conditions required for the spacetime to remain regular by analyzing the behavior of the metric and the Ricci scalar near the origin.
In particular, we find that the spacetime is regular and satisfies the energy conditions \footnote{Except for the strong energy condition, which is violated near the origin of a regular black hole.} when the parameters lie in the range
\begin{equation}
\frac{1}{D-2}<\omega_{1}\leq 1,
\qquad n>1.
\end{equation}
We also investigated the dynamical formation of these regular black holes using the thin-shell formalism. We derive the general evolution equation for a thin shell in arbitrary spacetime dimensions, assuming a linear barotropic equation of state. We find that for
\begin{equation}
\zeta = -\frac{D-3}{D-2},
\end{equation}
the gravitational collapse can produce a regular black hole. Depending on the initial conditions, the shell either collapses to form a regular black hole or reaches a finite minimum radius and then expands again, avoiding the formation of a spacetime singularity. We illustrate the evolution of the shell radius for both $D=4$, $D=5$ and $D=6$.

Overall, our results provide a simple and general framework for constructing regular black holes in higher dimensions from physically motivated matter sources and for studying their dynamical formation through gravitational collapse. This framework may also serve as a starting point for future investigations of the thermodynamics, stability, and observational properties of regular black holes in higher-dimensional and modified theories of gravity.

We also investigate the thermodynamics of the asymptotically AdS regular black holes in arbitrary dimensions. We verify that the solutions satisfy the first law of black hole thermodynamics and the corresponding Smarr relation. However, the Smarr formula is not equivalent to the Euler identity. We leave a detailed investigation of the relation between the Smarr formula and the Euler identity for future work.

Furthermore, we analyze the thermodynamic stability and phase structure of the black holes in four, five, and six spacetime dimensions. Our results show that these black holes exhibit dimension-dependent phase transitions, highlighting the important role of the spacetime dimension in determining their thermodynamic behavior.

These results provide a general framework for constructing regular AdS black holes in higher-dimensional Einstein gravity and for understanding both their dynamical formation and thermodynamic properties. Future work may include extending this analysis to modified theories of gravity.

\section*{Acknowledgements}
This research project is supported by the Second Century Fund (C2F).

\appendix
\section{The first law of thermodynamics and Smarr formula}\label{appsmarr}
To derive the variation of the mass function, let us define \cite{Ma:2014qma}
\begin{equation}
I(r_h,C_1,\rho_0)=\int_{r_h}^{\infty}f(r,C_1,\rho_0)\,dr.
\end{equation}
Since the integral depends on the horizon radius through the lower integration limit and on the matter parameters $C_1$ and $\rho_0$ through the integrand, its total variation receives contributions from all of these quantities. 
We first consider the variation of the lower integration limit i.e, $r_h\rightarrow r_h+\delta r_h.$ while keeping the integrand fixed. Then the integral becomes
\begin{equation}
I(r_h+\delta r_h)=\int_{r_h+\delta r_h}^{\infty}f(r,C_1,\rho_0)\,dr.
\end{equation}
Subtracting the original integral gives
\begin{equation}
\delta I_{\rm lim}=\int_{r_h+\delta r_h}^{\infty}f(r,C_1,\rho_0)\,dr-\int_{r_h}^{\infty}
f(r,C_1,\rho_0)\,dr.
\end{equation}
Using the identity
\begin{equation}
\int_{r_h+\delta r_h}^{\infty}=\int_{r_h}^{\infty}-\int_{r_h}^{r_h+\delta r_h},
\end{equation}
we obtain
\begin{equation}
\delta I_{\rm lim}=-\int_{r_h}^{r_h+\delta r_h}f(r,C_1,\rho_0)\,dr.
\end{equation}
Since $\delta r_h$ is infinitesimal, the integrand may be evaluated at the lower limit to first order,
\begin{equation}
\int_{r_h}^{r_h+\delta r_h}
f(r,C_1,\rho_0)\,dr=f(r_h,C_1,\rho_0)\,\delta r_h
+\mathcal{O}(\delta r_h^2).
\end{equation}
Therefore, we obtain
\begin{align}
\delta I_{\rm lim} &=-f(r_h,C_1,\rho_0)\,\delta r_h, \\
 &=-\Omega_{D-2}r_h^{D-2}T^{0}{}_{0}(r_h,C_1,\rho_0)\,\delta r_h.
\end{align}
Applying the Leibniz rule yields
\begin{align}
\dfrac{\delta I_{\rm lim}}{\Omega_{D-2}}={}&-r_h^{D-2}T^{0}{}_{0}\,\delta r_h+\left(\int_{r_h}^{\infty}r^{D-2}
\frac{\partial T^{0}{}_{0}}{\partial C_1}\,dr\right)\delta C_1+\left(\int_{r_h}^{\infty}r^{D-2}
\frac{\partial T^{0}{}_{0}}{\partial \rho_0}\,dr\right)\delta\rho_0.
\end{align}
Therefore, the variation of the gravitational mass
\begin{equation}
M=\frac{(D-2)\Omega_{D-2}}{16\pi}\left(r_h^{D-3}+\frac{r_h^{D-1}}{\ell^2}\right)-\Omega_{D-2}
\int_{r_h}^{\infty}r^{D-2}T^{0}{}_{0}\,dr,
\end{equation}
becomes
\begin{align}
\delta M={}&\Bigg[\frac{D-3}{4\pi r_h}+\frac{4Pr_h}{D-2}-\frac{4r_h\rho(r_h)}{D-2}\Bigg]\delta S
+V\,\delta P
\nonumber\\
&
-\Omega_{D-2}\left(\int_{r_h}^{\infty}r^{D-2}\frac{\partial \rho}{\partial C_1}\,dr\right)\delta C_1
-\Omega_{D-2}\left(\int_{r_h}^{\infty}r^{D-2}\frac{\partial\rho}{\partial \rho_0}\,dr
\right)\delta\rho_0.
\end{align}
On the other hand, one can obtain $C_{1}$ from \eqref{eqmass} as follows

\begin{equation}
C_{1}=\rho_{0}^{-(n-1)}\left[\frac{\Omega_{D-2}\rho_{0}}{M\alpha}B\!\left(\frac{D-1}{\alpha},\frac{1}{n-1}-\frac{D-1}{\alpha}\right)\right]^{\frac{\alpha}{D-1}},
\end{equation}
and the variation of $C_{1}(\rho_{0},M)$ is given by

\begin{equation}
\delta C_1=C_1\left[\left(\frac{\alpha}{D-1}-(n-1)\right)\frac{\delta\rho_0}{\rho_0}-\frac{\alpha}{D-1}
\frac{\delta M}{M}\right].
\end{equation}
Substituting $\delta C_{1}$ into the mass variation and collecting the $\delta M$ gives
\begin{align}\label{firstlaw}
&\Bigg[1+\frac{\alpha\,\Omega_{D-2}C_1}{(D-1)M}\left(\int_{r_h}^{\infty}r^{D-2}\frac{\partial\rho}{\partial C_1}\,dr\right)\Bigg]\delta M={}\Bigg[\frac{D-3}{4\pi r_h}+\frac{4Pr_h}{D-2}-\frac{4r_h\rho(r_h)}{D-2}\Bigg]\delta S+V\,\delta P
\nonumber\\
&
-\Omega_{D-2}\Bigg[\frac{C_1}{\rho_0}\left(\frac{\alpha}{D-1}-(n-1)\right)\left(\int_{r_h}^{\infty}
r^{D-2}\frac{\partial\rho}{\partial C_1}\,dr\right)+\left(\int_{r_h}^{\infty}r^{D-2}\frac{\partial\rho}{\partial\rho_0}\,dr\right)\Bigg]\delta\rho_0.
\end{align}
This can be rewritten as
\begin{equation}
    \mathcal{A}\delta M=T\delta S+V\delta P+\Psi_{\rho_{0}}\delta \rho_{0}.
\end{equation}
\\

The Smarr relation can be derived from the Komar integral identity rather than from Euler's theorem, since the physical mass originates entirely from the matter distribution. Starting from the Einstein equations \cite{Ma:2014qma,Balart:2017dzt,Rasheed:1997ns}
\begin{equation}
R_{\mu\nu}=8\pi\left(T_{\mu\nu}-\frac{1}{D-2}Tg_{\mu\nu}\right)+\frac{2\Lambda}{D-2}g_{\mu\nu},
\end{equation}
and using the timelike Killing vector $\xi^\mu=(1,0,\ldots,0)$ together with the Komar identity
\begin{equation}
\nabla_\mu\nabla^\mu\xi^\nu=-R^\nu{}_\mu\xi^\mu.
\end{equation}
One obtains the generalized Smarr relation
\begin{equation}
(D-3)M=(D-2)TS-2PV+\Delta,
\label{Smarr}
\end{equation}
where the matter contribution is
\begin{equation}
\Delta=16\pi\Omega_{D-2}\int_{r_h}^{\infty}r^{D-2}\left(T^{t}{}_{t}-\frac{T}{D-2}\right)dr.
\label{Delta}
\end{equation}
For the anisotropic fluid, the energy-momentum tensor is given by
\begin{equation}
T^\mu{}_\nu=\mathrm{diag}(-\rho,-\rho,P_t,\ldots,P_t),
\end{equation}
and its trace is
\begin{equation}
T=-2\rho+(D-2)P_t.
\end{equation}
Hence, we get
\begin{equation}
T^{t}{}_{t}-\frac{T}{D-2}=-\left(w_{1}+\frac{D-4}{D-2}\right)\rho-\omega_{2}\frac{\rho^{\,n}}{\rho_{0}^{\,n-1}},
\end{equation}
where
\begin{equation}
P_t=w_{1}\rho+\omega_{2}\frac{\rho^{\,n}}{\rho_{0}^{\,n-1}}.
\end{equation}
Therefore, the matter contribution becomes the following:
\begin{equation}
\Delta=-16\pi\Omega_{D-2}\int_{r_h}^{\infty}r^{D-2}\left[\left(w_{1}+\frac{D-4}{D-2}
\right)\rho(r)+\omega_{2}\frac{\rho(r)^n}{\rho_{0}^{\,n-1}}\right]dr.
\end{equation}
The mass contribution can be evaluated analytically. By defining, \footnote{In deriving the contribution from the upper integration limit, we have used the asymptotic expansion of the Gauss hypergeometric function; see, for example, Ref.~\cite{DLMF}, Sec.~15.12, or Ref.~\cite{GradshteynRyzhik}:
\begin{equation}
{}_2F_1(a,b;c;-x) \sim \frac{\Gamma(c)\Gamma(a-b)}{\Gamma(a)\Gamma(c-b)}
x^{-b},\qquad x\rightarrow\infty,
\end{equation}
valid for $\Re(a)>\Re(b)$. Setting $c=b+1$ gives
\begin{equation}
\lim_{x\to\infty}x^{\nu b}\,{}_2F_1(a,b;b+1;-x^\nu)
=b\,\frac{\Gamma(b)\Gamma(a-b)}{\Gamma(a)}.
\end{equation}}

\begin{equation}
A=\omega_{1}+\frac{D-4}{D-2},
\qquad
B=C_{1}\rho_{0}^{\,n-1},
\end{equation}
one finds
\begin{align}
\mathcal{I}&=\frac{\rho_{0}}{\alpha}B^{-\frac{D-1}{\alpha}}\Gamma\!\left(\frac{D-1}{\alpha}\right)
\nonumber\\
&\times
\left[A\frac{\Gamma\!\left(\dfrac{\omega_{1}(D-2)-1}{(\omega_{1}+1)(n-1)(D-2)}\right)}{\Gamma\!\left(\dfrac1{n-1}\right)}
+\omega_{2}\frac{\Gamma\!\left(\dfrac{n(\omega_{1}+1)(D-2)-(D-1)}{(\omega_{1}+1)(n-1)(D-2)}\right)}{\Gamma\!\left(\dfrac{n}{n-1}\right)}\right]
\nonumber\\
&\quad
-\frac{\rho_{0}r_h^{D-1}}{D-1}\Bigg[A\,{}_2F_1\!\left(\frac1{n-1},\frac{D-1}{\alpha};1+\frac{D-1}{\alpha};-B r_h^{\alpha}
\right)
\nonumber\\
&\hspace{4.5cm}
+\omega_{2}\,{}_2F_1\!\left(\frac{n}{n-1},\frac{D-1}{\alpha};1+\frac{D-1}{\alpha};-B r_h^{\alpha}\right)\Bigg],
\label{eq:pressureintegral}
\end{align}
and $\Delta$ is given by
\begin{equation}\label{eqdelta}
\Delta =-16\pi\Omega_{D-2}\mathcal{I}.
\end{equation}


\bibliographystyle{unsrt}
\bibliography{references}


\end{document}